\documentclass[12pt]{article}
\usepackage{amsmath,amssymb,exscale,cancel}
\usepackage{slashed}
\allowdisplaybreaks
\usepackage{graphicx}
\usepackage{graphicx}
\usepackage{epsfig}
\usepackage{multicol}
\usepackage{color}
\usepackage{mathrsfs}
\usepackage{blindtext}
 \usepackage{fancyhdr}
\usepackage{hyperref}
\usepackage{cite}
\usepackage{mathtools}

\usepackage{floatrow}

\usepackage{graphicx}
\usepackage{sidecap}

\usepackage[latin1]{inputenc} 

\usepackage{multicol}  
 
 \usepackage{titlesec}
 
 \usepackage{rotating,slashed,xcolor,amsfonts,expdlist,charter}

\numberwithin{equation}{section}

\usepackage{xcolor}
\usepackage{sectsty}

\usepackage{mdframed}
\usepackage{titletoc}

\numberwithin{equation}{section}

\usepackage{xcolor}
\usepackage{sectsty}

\usepackage{mdframed}
\usepackage{titletoc}

\definecolor{secnum}{RGB}{13,151,225}
\definecolor{ptcbackground}{RGB}{212,237,252}
\definecolor{ptctitle}{RGB}{0,177,235}

\titlecontents{lsection}
  [5.8em]{\sffamily}
  {\color{secnum}\contentslabel{2.3em}\normalcolor}{}
  {\titlerule*[1000pc]{.}\contentspage\\\hspace*{-5.8em}\vspace*{5pt}%
    \color{white}\rule{\dimexpr\textwidth-15.5pt\relax}{1pt}}

\usepackage{hyperref}
\hypersetup{colorlinks,bookmarksopen,bookmarksnumbered,citecolor=rossos,
linkcolor=redy,pdfstartview=FitH,urlcolor=green-go}
\usepackage{slashed}

\definecolor{blus}{cmyk}{1,0.9,0,0.1}
\definecolor{verdes}{cmyk}{0.99,0,0.59,0.65}
\definecolor{rossos}{cmyk}{0,1,1,0.55}
\definecolor{redy}{cmyk}{0,1,1,0.7}
\definecolor{greeny}{cmyk}{0.99,0,0.59,0.98}
\definecolor{green-go}{cmyk}{0.79,0,0.59,0.5}

\usepackage{titlesec}

\def\Lag{\mathscr{L}}

\newcommand{\beq}{\begin{equation}}
\newcommand{\eeq}{\end{equation}}

\def\hhref#1{\href{http://arxiv.org/abs/#1}{arXiv:#1}} 

 \def\Lag{\mathscr{L}}
 
\newcommand{\tmtextbf}[1]{{\bfseries{#1}}}
\newcommand{\tmtextrm}[1]{{\rmfamily{#1}}}
\newcommand{\bp}{\bar M_P}

\def\be{\begin{equation}}
\def\ee{\end{equation}}
\def\ba{\begin{array} }

\def\bac{\begin{array} {c}}
\def\bacc{\begin{array} {cc}}
\def\baccc{\begin{array} {ccc}}
\def\bacccc{\begin{array} {cccc}}
\def\ea{\end{array}}
\def\bea{\begin{eqnarray}}
\def\eea{\end{eqnarray}}

\definecolor{red}{rgb}{1,0,0}

\def\psl{\hbox{\hbox{${p}$}}\kern-1.9mm{\hbox{${/}$}}}
\def\dsl{\hbox{\hbox{${\partial}$}}\kern-2.2mm{\hbox{${/}$}}}
\def\Dsl{\hbox{\hbox{${D}$}}\kern-2.6mm{\hbox{${/}$}}}

\def\Lag{\mathscr{L}}

\newcommand{\gappeq}{{\rlap{{\raise}.5ex\text{\ensuremath{>}}}{{\lower}.5ex\text{\ensuremath{\sim}}}}}
\newcommand{\lappeq}{{\rlap{{\raise}.5ex\text{\ensuremath{<}}}{{\lower}.5ex\text{\ensuremath{\sim}}}}}
\newcommand{\I}{\tmtextrm{1{\kern}-.24em l}}

\begin{document}
\topmargin -1.0cm
\oddsidemargin 0.9cm
\evensidemargin -0.5cm

{\vspace{-1cm}}
\begin{center}

\vspace{-1cm}

 {\tmtextbf{ 
 \hspace{-1.2cm}   
{\LARGE  \color{rossos}Supercooled Phase Transitions  \\ with Radiative Symmetry Breaking}
 }} {\vspace{.5cm}}\\

\vspace{1.3cm}

{\large{\bf  Alberto Salvio }}

{\em  
\vspace{.4cm}
 Physics Department, University of Rome Tor Vergata, \\ 
via della Ricerca Scientifica, I-00133 Rome, Italy\\

\vspace{0.6cm}

I. N. F. N. -  Rome Tor Vergata,\\
via della Ricerca Scientifica, I-00133 Rome, Italy\\

  \vspace{0.5cm}

}
\vspace{1.5cm}
\end{center}

\noindent ---------------------------------------------------------------------------------------------------------------------------------
\begin{center}
{\bf \large Abstract}
\end{center}
\noindent   First-order phase transitions produce gravitational waves and primordial black holes. They always occur in field theories where symmetries are radiatively broken and masses are correspondingly generated. These theories predict a period of supercooling: phase transitions become effective at temperatures much smaller than the symmetry-breaking scale. This paper reviews a model-independent approach to study phase transitions in this scenario, which can be adopted if supercooling is strong enough. Perturbative methods can be used to determine the effective action and such model-independent approach allows us to obtain ready-to-use formulas that can be applied to any specific model of this sort.

\vspace{0.7cm}

\noindent---------------------------------------------------------------------------------------------------------------------------------

\newpage

\tableofcontents

\noindent --------------------------------------------------------------------------------------------------------------------------------

\vspace{0.2cm}

\section{Introduction}\label{intro}

The interplay between particle physics and cosmology offers the possibility of testing fundamental theories at energies much higher than those accessible at particle accelerators.  A classic example is inflation, which can involve energies a few orders of magnitude below the Planck scale. 

Another example is furnished by first-order cosmological phase transitions (PTs), which represent a violent process because of bubble nucleation. These PTs can (although they do not have to) be associated with a symmetry-breaking scale several orders of magnitude above the electroweak (EW) scale. Furthermore, they can produce a background of gravitational waves (GWs) that may be observable (see Ref.~\cite{Maggiore:2018sht} for a textbook introduction) and primordial black holes (PBHs)~\cite{Hawking:1982ga,Crawford:1982yz,Kodama:1982sf,Moss:1994iq,Khlopov:1999ys,Khlopov:2008qy,Gross:2021qgx,Kawana:2021tde,Liu:2021svg,Baker:2021sno,Jung:2021mku,Hashino:2021qoq,Hashino:2022tcs,Kawana:2022olo,Lewicki:2023ioy,Gouttenoire:2023naa,Gouttenoire:2023bqy,Salvio:2023ext,Gouttenoire:2023pxh,Salvio:2023blb,Conaci:2024tlc,Lewicki:2024ghw,Balaji:2024rvo,Cai:2024nln,Ai:2024cka,Arteaga:2024vde,Goncalves:2024vkj,Banerjee:2024fam,Zhang:2025kbu,Ning:2026nfs}, which can account for (a fraction of) dark matter. In the last ten years interest in GW astronomy and compact objects, such as (primordial) black holes, has significantly increased following the discovery of the GWs from  binary black hole and neutron star mergers~\cite{Abbott:2016blz,TheLIGOScientific:2016wyq1,LIGOScientific:2017ync}. Recently, the interest in this field has been further boosted by the detection of a background of GWs by pulsar timing arrays, including the North American Nanohertz Observatory for Gravitational Waves (NANOGrav), the Chinese Pulsar Timing Array (CPTA), the European Pulsar Timing Array (EPTA) and the Parkes Pulsar Timing Array (PPTA)~\cite{NANOGrav:2023gor,Antoniadis:2023ott,Reardon:2023gzh,Xu:2023wog}.

It is interesting to note that the possible observation of effects generated by cosmological first-order PTs (FOPTs) would be a clear signal of new physics, because the SM does not feature this type of PT in a cosmological setting.

In general, it is not possible to perturbatively describe FOPTs in models where the corresponding symmetry breaking arises entirely from the Higgs mechanism because of the presence of complex effective potentials~\cite{Salvio:2020axm,Salvio:2023qgb,Salvio:2024upo} and other issues at high-temperatures~\cite{Weinberg:1974hy,Linde:1980ts,Linde:1978px}. These problems can be avoided in theories where  the corresponding symmetry breaking is mostly induced (and masses are then generated) radiatively, i.e.~through perturbative loop corrections\footnote{More generally, it is not necessary for the entire theory to rely on this mechanism; it is enough that a sector of the theory with sufficiently small couplings to the other sectors  exhibits this behavior.}~\cite{Salvio:2023qgb,Salvio:2024upo}. Moreover, when this radiative symmetry breaking (RSB) occurs, the corresponding PT is always of first order. The seminal work on RSB is Ref.~\cite{Coleman:1973jx} by Coleman and E.~Weinberg, who studied a simple toy model (see also Ref.~\cite{Levi:2022bzt}
for a subsequent analysis). The Coleman-Weinberg work was then extended to broader field theories by Gildener and S.~Weinberg~\cite{Gildener:1976ih}. Furthermore, thanks to an approximate scale invariance, in the RSB scenario the FOPTs feature a period of supercooling, when the temperature $T$ drops by several orders of magnitude below the critical temperature~\cite{Witten:1980ez,Salvio:2023qgb}.
Supercooling is one of the key properties of the RSB scenario that validates the one-loop perturbative approximation (as well as the derivative expansion). It is also interesting to note that during supercooling the universe expands exponentially, leading to another inflationary period.

Another motivation of theories with RSB is the fact that the dimensionful quantities can be generated through dimensional transmutation and this effect can  generically lead to exponentially large hierarchies between mass scales~\cite{Gildener:1976ih}. This is because the renormalization group (RG) running of a generic dimensionless coupling $\kappa$ is logarithmic in the RG energy scale $\mu$. Therefore, order-one variations of $\kappa$ correspond to exponentially large variations of $\mu$. This is the reason why physicists do not consider the large hierarchy between the QCD confining scale and the Planck mass to be puzzling, but rather see it as a quite natural fact. 
\vspace{0.1cm}

{\it This paper reviews a model-independent approach to describe FOPTs in RSB theories with sufficiently high supercooling; the latter hypothesis  leads to an expansion in terms of a  quantity that is small when supercooling is large enough.} 
\vspace{0.1cm}

This review paper is based on an invited talk by the author at the AstroMarche2 Conference (New Avenues for Classical and Quantum Cosmology), which took place in the Italian city of Camerino in September 2025. In turn, that talk was  mainly based on Refs.~\cite{Salvio:2023qgb,Salvio:2023ext,Salvio:2023blb,Banerjee:2024cwv,Rescigno:2025ong}. A detailed summary of the topics covered here is given in the concluding Sec.~\ref{Conclusions}.

\section{General theoretical framework}\label{General theoretical framework}

  In this section we discuss the general class of theories where supercooled PTs occur and whose effective action can be studied with perturbative methods: the  RSB scenario. Furthermore, we will see how a period of supercooling takes place. In the subsequent sections supercooling will be a crucial ingredient of the model-independent approach to PTs.

  \subsection{Fields and Lagrangian}\label{Fields and Lagrangian}

Since in the RSB scenario the mass spectrum predominantly emerges from loop-induced effects rather than tree-level terms, we start from the most general no-scale matter Lagrangian describing  the interactions between the matter fields\footnote{Repeated indices understand a summation.}:
\be \label{eq:Lmatterns}
\Lag^{\rm ns}_{\rm matter} =  
- \frac14 F_{\mu\nu}^AF^{A\mu\nu} + \frac{D_\mu \phi_a \, D^\mu \phi_a}{2}  + \bar\psi_j i\slashed{D} \psi_j  - \frac12 (Y^a_{ij} \psi_i\psi_j \phi_a + \hbox{h.c.}) 
- V_{\rm ns}(\phi), 
\ee 
while gravity is assumed to be described by standard Einstein's theory at the energies that are relevant for this work\footnote{However, it is possible to construct a classically scale-invariant theory of gravity where scale invariance is broken by quantum effects~\cite{Salvio:2014soa,Kannike:2015apa,Kannike:2015fom,Salvio:2015wka,Salvio:2017qkx,Salvio:2017xul,Salvio:2019wcp,Alvarez-Luna:2022hka,Cecchini:2024xoq} at energy scales that are assumed to be above those of interest here (see~\cite{Salvio:2020axm} for a review).}.
Here we consider generic numbers 
of real scalars $\phi_a$,   Weyl fermions $\psi_j$ and vectors $V^A_\mu$ (with field strength $F_{\mu\nu}^A$), so we can cover field theories of phenomenological relevance. The gauge fields $V^A_\mu$
allow us to construct the covariant derivatives
$$D_\mu \phi_a = \partial_\mu \phi_a+ i \theta^A_{ab} V^A_\mu \phi_b, \qquad D_\mu\psi_j = \partial_\mu \psi_j + i t^A_{jk}V^A_\mu\psi_k, $$ 
where the $\theta^A$ and $t^A$ are the generators of the internal gauge group in the scalar and fermion representations, respectively. They are Hermitian matrices and contain the gauge couplings. Note that, since we are working with real scalars, the  $\theta_A$ are purely imaginary and antisymmetric. 
Also, the $Y^a_{ij}$  are the Yukawa couplings  and $V_{\rm ns}(\phi)$ is the no-scale potential,
\be V_{\rm ns}(\phi)= \frac{\lambda_{abcd}}{4!} \phi_a\phi_b\phi_c\phi_d, \label{Vns}\ee
where the $\lambda_{abcd}$ are the quartic couplings. We take $\lambda_{abcd}$ totally symmetric with respect to the exchange of its indices $abcd$ without loss of generality. In~(\ref{eq:Lmatterns}) all terms must respect gauge invariance.

Since dimensionful coefficients do not appear in the classical Lagrangian, these theories have a great deal of predictive power, see~\cite{Gildener:1976ih} for a general discussion and e.g.~\cite{Ghoshal:2020vud} for a specific example.
 
 Models belonging to this class may explain the background of GWs recently detected by pulsar timing arrays~\cite{Salvio:2023blb,Balan:2025uke,Bringmann:2026xcx}.
 
 \subsection{Radiative symmetry breaking}\label{CWGW}

  In the RSB mechanism the mass scales emerge radiatively from loops as follows~\cite{Coleman:1973jx,Gildener:1976ih}.
   The basic idea is that, since at quantum level the couplings depend on the RG energy $\mu$, 
   at a particular RG energy, the potential in Eq.~(\ref{Vns}) (with running  $\lambda_{abcd}$) can exhibit a flat direction in field space.
Such flat direction can be written as $\phi = \nu \chi$, where $\nu$ is a unit vector, i.e.~$ \nu_a  \nu_a =1$, and $\chi$ is a single real scalar field, which parameterizes this direction.  
Therefore, after renormalization, the RG-improved potential $V$ along the flat direction reads
\be V(\chi) = \frac{\lambda_\chi (\mu)}{4}\chi^4,  \label{Vvarphi}\ee 
where 
\be \lambda_\chi(\mu) \equiv\frac1{3!} \lambda_{abcd}(\mu) \nu_a \nu_b \nu_c \nu_d. \label{lambdaphi}\ee
Having a flat direction along $\nu$ for $\mu$ equal to some specific value $\tilde\mu$ means 
 \be \lambda_\chi(\tilde\mu)\equiv\lambda_{abcd}(\tilde\mu) \nu_a \nu_b \nu_c \nu_d=0. \label{FlatCond}\ee
 
 It is precisely the existence of a flat (or quasi-flat) direction that allows to break symmetries and generate masses only radiatively in a perturbative theory. 
Indeed, in the absence of such flat direction (or approximate flat direction) it would not be possible to describe the minimum of the potential within perturbation theory~\cite{Gildener:1976ih}. We will revisit and expand this argument below in this section (see the discussion around Eq.~(\ref{NPchi})). It is for this reason that letting the renormalization scale acquire values different from $\tilde\mu$, as done for example in~\cite{Pascoli:2026tuu}, may result in a breakdown rather than an improvement of perturbation theory. Thus couplings should remain small enough to keep loop corrections and logarithms under perturbative limits.

  Besides the potential in~(\ref{Vvarphi}), quantum loop corrections also generate other terms $V_1+V_2+...$, where $V_i$ represents the $i$th-loop contribution. The explicit expression of $V_1$ is well known. One can recover it, without specifying the details of the underlying theory, by recalling that the effective potential does not depend on $\mu$. Indeed, the renormalization changes the couplings, the masses and  the fields, but leaves the Lagrangian (and in particular the potential) unchanged. In formul\ae
 \be \mu \frac{dV_q}{d\mu} =0, \qquad \mbox{where}\quad V_q\equiv V+V_1+V_2+...\, .\ee
 Using~(\ref{Vvarphi}), the solution of this equation at the one-loop level is 
 \be V_q=\frac{\lambda_\chi (\mu)}{4}\chi^4+ \frac{\beta_{\lambda_\chi}}4\left(\log\frac{\chi}{\mu}+a_s\right)\chi^4, \label{Vqs} \ee
 where 
 \be \beta_{\lambda_\chi} \equiv  \mu\frac{d\lambda_{\chi}}{d\mu} \ee
 is the beta function of $\lambda_\chi$ 
 and  $a_s$ is a renormalization-scheme-dependent quantity. Setting $\mu=\tilde\mu$, where $\lambda_\chi=0$, leads to
 \be V_q(\chi) = \frac{\bar \beta}4\left(\log\frac{\chi}{\chi_0}-\frac14\right)\chi^4,\label{CWpot}\ee
where
 \be \bar\beta \equiv \left[\beta_{\lambda_\chi} \right]_{\mu=\tilde\mu}, \qquad \chi_0\equiv \frac{\tilde\mu}{e^{1/4+a_s}}. \label{barbetachi0}\ee
 Note that the renormalization-scheme-dependent $a_s$ has been absorbed in the scale $\chi_0$.
 We see that the flat direction acquires some steepness radiatively.  The field value $\chi_0$ is a stationary point of $V_q$.
Moreover, $\chi_0$ is a point of minimum if $\bar\beta>0$. Therefore, when the conditions 
  \beq\left\{
\begin{array}{rcll}
\lambda_\chi(\tilde\mu)  &=& 0 & \hbox{(flat direction),}\\
 & & \\ 
\beta_{\lambda_\chi}(\tilde\mu)  &>& 0 & \hbox{(minimum condition),}
\end{array}\right.
\label{eq:CWgen}
\eeq
 are satisfied quantum corrections generate a minimum of the potential at a non-vanishing value of $\chi$, i.e.~$\chi_0$. In that case $\chi_0$ is the (radiatively induced) zero-temperature vacuum expectation value of $\chi$ and the fluctuations of $\chi$ around $\chi_0$ have mass 
 \be m_\chi=\sqrt{\bar\beta}\, \chi_0, \label{mchi} \ee
 which is real and positive from the second condition in~(\ref{eq:CWgen}). 
  
  If we had chosen another value of $\mu$, with a sizable $\lambda_\chi$, we would not have been able to describe the minimum of $V_q$ within perturbation theory: formally~(\ref{Vqs}) would have been stationary at the non-perturbative value of $\chi$ such that 
 \be\log\frac{\chi}{\mu} = -\frac{\lambda_\chi}{\beta_{\lambda_\chi}}-a_s-\frac14.\label{NPchi}\ee
   The corresponding value of $\chi$ would have been non perturbative because  $\beta_{\lambda_\chi}$ is at least quadratic in the coupling constants and the ratio $\frac{\lambda_\chi}{\beta_{\lambda_\chi}}$ appears in~(\ref{NPchi}), implying that some positive powers of coupling constants appear in the denominator. As anticipated above, we see the importance of setting $\mu=\tilde \mu$, where the flat-direction condition in~(\ref{FlatCond}) holds. This argument also shows to what extent a quasi-flat direction is acceptable: since in the right-hand side of~(\ref{NPchi}) the ratio $\lambda_\chi/\beta_{\lambda_\chi}$ appears, the condition on $\mu$ such that perturbation theory is preserved is that the corresponding value of $\lambda_\chi/\beta_{\lambda_\chi}$ remains sufficiently small. This is a rather strong condition because $\beta_{\lambda_\chi}$ is not only suppressed by some powers of coupling constants, but also by small loop factors of order at most $1/(4\pi)^2$, if perturbation theory needs to be preserved.
  
  The non-trivial minimum, $\chi_0$, can generically break global and/or gauge symmetries and thus generate the  particle masses, with $\chi_0$ playing the role of the symmetry-breaking scale.  Consider, for example, a term in the Lagrangian density $\Lag$ of the form 
 \be \Lag_{\chi h}\equiv \frac12 \lambda_{ab} \phi_a\phi_b |{\mathcal H}|^2,\label{LvarphiH}\ee 
 where ${\mathcal H}$ is the Standard Model (SM) Higgs doublet and the $\lambda_{ab}$ are some of the quartic couplings. RG-improving and setting $\mu=\tilde\mu$ and $\phi$ along the flat direction, $\nu$,
 \be  \Lag_{\chi h} = \frac12 \lambda_{\chi h}(\tilde\mu) \chi^2 |{\mathcal H}|^2,\label{portal} \ee
 where 
 \be \lambda_{\chi h}(\mu) \equiv  \lambda_{ab}(\mu) \nu_a\nu_b.\ee
 Thus, by evaluating this term at the minimum where $\chi=\chi_0$ we obtain the Higgs squared mass parameter
  \be \mu_h^2 \equiv \frac12\lambda _{\chi h}(\tilde\mu) \chi_0^2. \label{muh}\ee
 In order to provide a mass to the SM elementary particles, one can require $\mu_h^2>0$, namely we have the additional condition
 \be \lambda _{\chi h}(\tilde\mu) >0\quad  \hbox{(generation of the EW scale)}. \ee

Recalling the well-known formula that relates $\mu_h^2$ and the physical Higgs mass $M_h$, one finds the condition $\sqrt{\lambda_{\chi h}}\chi_0\approx M_h\approx 125$~GeV.
So we cannot use this mechanism to generate $\mu_h^2$ when $\chi_0$ is much below the EW scale. This is  due both to collider limits on extra light scalars with strong coupling to the Higgs and  the requirement of validity of perturbation theory. Of course, it is still possible that the SM with an explicit scale-symmetry breaking parameter is coupled to an approximately scale-invariant sector that features RSB. In this case perturbation theory can be compatible with a $\chi_0$ much smaller than the EW scale and such scale-invariant sector must be ``dark", i.e. must have only tiny couplings with the SM particles (see~\cite{Salvio:2023blb} for an example).

\subsection{Thermal effective potential}\label{TVeff}

 In order to write a model-independent formula for the thermal contribution to the effective potential, $V_{\rm eff}$, we need to write general expressions for the background-dependent masses.

In the scalar sector the squared-mass matrix is given by the Hessian matrix  of the no-scale classical potential in~(\ref{Vns}):
\be M_{Sab}^2  \equiv \frac{\partial^2V_{\rm ns}}{\partial\phi_a\partial\phi_b} =\frac12 \lambda_{abcd} \phi_c\phi_d. \label{MS20}\ee 
By evaluating this Hessian matrix at the flat direction, $\phi =\nu\chi$, we obtain 
\be M_{Sab}^2(\chi)=\frac12\lambda_{abcd}\nu_c\nu_d\chi^2. \label{MS2}\ee
Since $M_S^2$ is real and symmetric it can be diagonalized with a real orthogonal matrix, to obtain $M_S^2(\chi)\to$\,\,diag$(...,m_s^2(\chi), ...)$, where the $m_s^2(\chi)$ are the background-dependent scalar squared  masses, the eigenvalues of $M_S^2(\chi)$. All the  $m_s^2$ must be non-negative to have a classical potential bounded from below. To prove this first note that 
the requirement that the classical potential in~(\ref{Vns}) be bounded from below also implies that the potential has the absolute minimum at $\phi=0$  and this minimum vanishes (because no scales are present in~(\ref{Vns})). Also the classical potential is constantly equal to its value at $\phi=0$ along the flat direction (by definition of flat direction). So for a classical potential that is bounded from below and has a flat direction $\nu\chi$
\be V_{\rm ns}(\phi) = V_{\rm ns}(\nu \chi)+\frac{\partial V_{\rm ns}}{\partial\phi_a}(\nu \chi) \delta\phi_a+\frac12\frac{\partial^2V_{\rm ns}}{\partial\phi_a\partial\phi_b}(\nu \chi) \delta\phi_a\delta\phi_b=\frac12\frac{\partial^2V_{\rm ns}}{\partial\phi_a\partial\phi_b}(\nu \chi) \delta\phi_a\delta\phi_b,  \label{ArgPosMS}\ee
where $\delta\phi\equiv \phi-\nu\chi$ is taken here infinitesimal and therefore the corrections of order of the cube of $\delta\phi$ are neglected. So if there were negative eigenvalues of $M_{Sab}^2(\chi)$ the classical potential would become smaller than its value at the flat direction, but we have seen that this is not possible for  a bounded-from-below potential. So all $m_s(\chi)$ must be real and, of course, can be taken non-negative.

 This nice property is not shared by theories where symmetry breaking is entirely due to the standard Higgs mechanism, which always requires non-convex regions of the tree-level potential and thus some negative scalar squared masses for some field values.  This pathology is a manifestation of the breaking of the perturbative (loop) expansion and occurs because the field configurations where the scalar potential is not convex are too far from the minima of the scalar action~\cite{Salvio:2024upo}.

Let us now turn to the fermion sector. For a given constant field background $\phi$, we choose a fermion basis where $\mu_F \equiv Y^a\phi_a$ (and so also $\mu_F^\dagger$) is diagonal, which can be obtained through an $SU(N_F)$ transformation acting on the fermion fields\footnote{This is known as the complex Autonne-Takagi factorization, see also Ref.~\cite{Youla}.}. The squared-mass matrix is
\be M_F^2 \equiv \mu_F\mu_F^\dagger. \ee 
By evaluating $\phi$ at the flat direction one obtains $\mu_F(\chi) = Y_\nu \chi$, where $Y_\nu\equiv Y^a\nu_a$, and 
\be M_F^2(\chi) =  Y_\nu Y_\nu^\dagger \chi^2. \ee  
In our fermion basis $M_F^2(\chi)=$\,\,diag$(...,m_f^2(\chi), ...)$, where the $m_f(\chi)$ are the background-dependent fermion masses.
Given that $M_F^2$ is the product of $\mu_F$ times its Hermitian conjugate all the $m_f$ are real and, of course, can be taken non-negative.

Finally, in the vector sector the elements of the squared-mass matrix $M^2_V$ are
\be M^2_{VAB}\equiv \phi^T\theta^A\theta^B\phi,\ee
and, evaluating at the flat direction, $M^2_{VAB}(\chi)\equiv \nu^T\theta^A\theta^B\nu \chi^2$.
Since the $\theta^A$ are Hermitian, purely imaginary and antisymmetric, $M^2_V$ is always real, symmetric and non-negatively defined: one can diagonalize $M^2_V$ with a real orthogonal matrix, to obtain $M^2_V\to$
\,\,diag$(...,m_v^2(\chi), ...)$, where the $m_v(\chi)$ are the background-dependent vector masses (the eigenvalues of $M_V^2(\chi)$) and are, just like the $m_s$ and $m_f$, all real. Of course, they can also be taken non-negative.

Including the thermal corrections, the general expression of the one-loop effective potential $V_{\rm eff}$ is then 
 (in the Landau gauge, see~\cite{Salvio:2024upo} for a pedagogical derivation)
\be V_{\rm eff}(\chi,T) = V_q(\chi) +\frac{T^4}{2\pi^2}\left(\sum_b n_b J_B(m_b^2(\chi)/T^2)-2\sum_f J_F(m_f^2(\chi)/T^2)\right)+\Lambda_0,  \label{VeffSumm}  \ee
where $V_q(\chi)$ is given in~(\ref{CWpot}), the sum over $b$ runs over all bosonic degrees of freedom and $n_b=1$ for a scalar degree of freedom (recall that we work with real scalars) and $n_b=3$ for a vector degree of freedom.  In~(\ref{VeffSumm}) the sum over $f$, which runs over the fermion degrees of freedom, is multiplied by 2 because we work with Weyl spinors. Also,  
the thermal functions $J_B$ and $J_F$ are defined by 
\bea \, \hspace{-1cm}J_B(x)\equiv \int_0^\infty dp\, p^2 \log\left(1-e^{-\sqrt{p^2+x}}\right)&=&-\frac{\pi^4}{45}+\frac{\pi^2}{12} x -\frac{\pi}{6} x^{3/2} -\frac{x^2}{32} \log\left(\frac{x}{a_B}\right) + O(x^3), \label{JBdef}\\
\hspace{-1cm}J_F(x)\equiv \int_0^\infty dp\, p^2 \log\left(1+e^{-\sqrt{p^2+x}}\right)&=& \frac{7\pi^4}{360}-\frac{\pi^2}{24} x -\frac{x^2}{32} \log\left(\frac{x}{a_F}\right) + O(x^3) \label{JFdef}.\eea
In the expressions above we also wrote the expansions of $J_B(x)$ and $J_F(x)$ around $x=0$ modulo terms of order
$x^3$, where $a_B = 16\pi^2 \exp(3/2-2\gamma_E)$, $a_F = \pi^2 \exp(3/2-2\gamma_E)$ and  $\gamma_E$ is the Euler-Mascheroni constant (see Ref.~\cite{Dolan:1973qd} for the derivation of those expansions). Moreover, all terms of the expansions of $J_B(x)$ and $J_F(x)$ around $x=0$ can be found, for example, in~\cite{Quiros:1994dr}. In Eq.~(\ref{VeffSumm}) we have included  a constant term $\Lambda_0$ to account for the observed value of the cosmological constant when $\chi$ is set to the point of minimum of $V_{\rm eff}$.
Adding $\Lambda_0$ does not spoil the argument presented above showing that for a scale-invariant and bounded-from-below potential with a flat direction the eigenvalues of $M_S^2$ are always non-negative: adding $\Lambda_0$ to $V_{\rm ns}$ just produces an additive constant $\Lambda_0$ on the right-hand side of Eq.~(\ref{ArgPosMS}) and $V_{\rm ns}$ would still  be unbounded from below if there were negative  eigenvalues of $M_S^2$.

It is important to note that supercooling (i.e.~$\chi_0\gg T$), to be discussed in Sec.~\ref{supercool}, guarantees that we are far from the high-temperature regime for which the perturbative expansion is known to break down~\cite{Weinberg:1974hy,Linde:1980ts,Linde:1978px}, so a one-loop computation should be a good approximation for small enough couplings~\cite{Salvio:2023qgb}. This is because the fields that participate in the transition (that directly interact with the flat-direction field $\chi$) receive, thanks to supercooling, a zero-temperature mass that is much larger than the thermal mass and the infrared problem discussed in~\cite{Linde:1980ts,Linde:1978px} is avoided. 

Moreover, resumming the daisy graphs~\cite{Arnold:1992rz} does not alter this conclusion. The reason is the following. In order to keep perturbation theory valid in this resummation one should include not only the thermally induced masses (which are also known as the thermal masses) as done, for example, in~\cite{Kierkla:2023von,Pascoli:2026tuu}, but also the radiatively generated ones, which are proportional to $\chi_0$.
Therefore, all the daisy-diagram contributions are suppressed by the  ratio $\chi_0/T$, which is large due to supercooling  ($T \ll \chi_0$). A way to see this explicitly is to organize the perturbative calculations based on these Feynman diagrams using the fields that represent the physical fluctuations around the true point of minimum of the potential, $\chi_0$. Indeed, this diagrammatic perturbative calculation should be organized  around a minimum of the potential. If one does so, the masses of the particles running in the loops should include the zero-temperature contribution due to $\chi_0$. Such contribution makes the Bose-Einstein and Fermi-Dirac distributions exponentially small in the presence of supercooling, which is always the case in the RSB scenario, as shown in Sec.~\ref{supercool}. So the corresponding thermal contributions are small in a RSB scenario. In the theories considered in~\cite{Arnold:1992rz}, on the other hand, a different situation is examined, in which the temperature is large compared to the zero-temperature mass.

Note that, from two loops on, Linde's problem discussed in~\cite{Linde:1980ts,Linde:1978px} can potentially reappear because there may be some unbroken gauge symmetry that indirectly couple to $\chi$ through messenger fields. For example these messenger fields can be some fermions that receive a mass from $\chi$ and are charged under the unbroken gauge symmetry. However, once the above-mentioned resummation  is performed, these higher-loop contributions are free from infrared divergences  and are suppressed compared to the one-loop contribution due to loop factors, which should be present for the general validity of the loop expansion.

Since in the RSB mechanism all $m_s^2$, as well as all $m_f^2$ and $m_v^2$, are non-negative the effective potential is guaranteed to be real. This is not the case in theories where symmetry breaking occurs through the standard Higgs mechanism: since the tree-level potential  is not always convex some of the $m_s^2$ are necessarily negative for some field values and the effective potential acquires an imaginary part there. 
This pathology is a manifestation of the breaking of the  (loop) perturbation theory: it occurs because the loop expansion is an expansion around the minima of the scalar action and the regions where the tree-level potential is not convex are too far from those minima. Therefore, as long as the couplings are small enough, the RSB mechanism supports the validity of perturbation theory in the calculation of the effective potential.  

\subsection{Radiatively-induced phase transitions are of first order}\label{PT}
  
  We are now ready to study the PTs associated with a RSB in our general theory~(\ref{eq:Lmatterns}). In  field theory the role of the order parameter can be played by the expectation value $\langle\chi\rangle$, which includes quantum as well as thermal averages: it is the point of minimum of the full effective potential, including both quantum and thermal contributions.

\begin{figure}[t]
\begin{center}
  \includegraphics[scale=0.5]{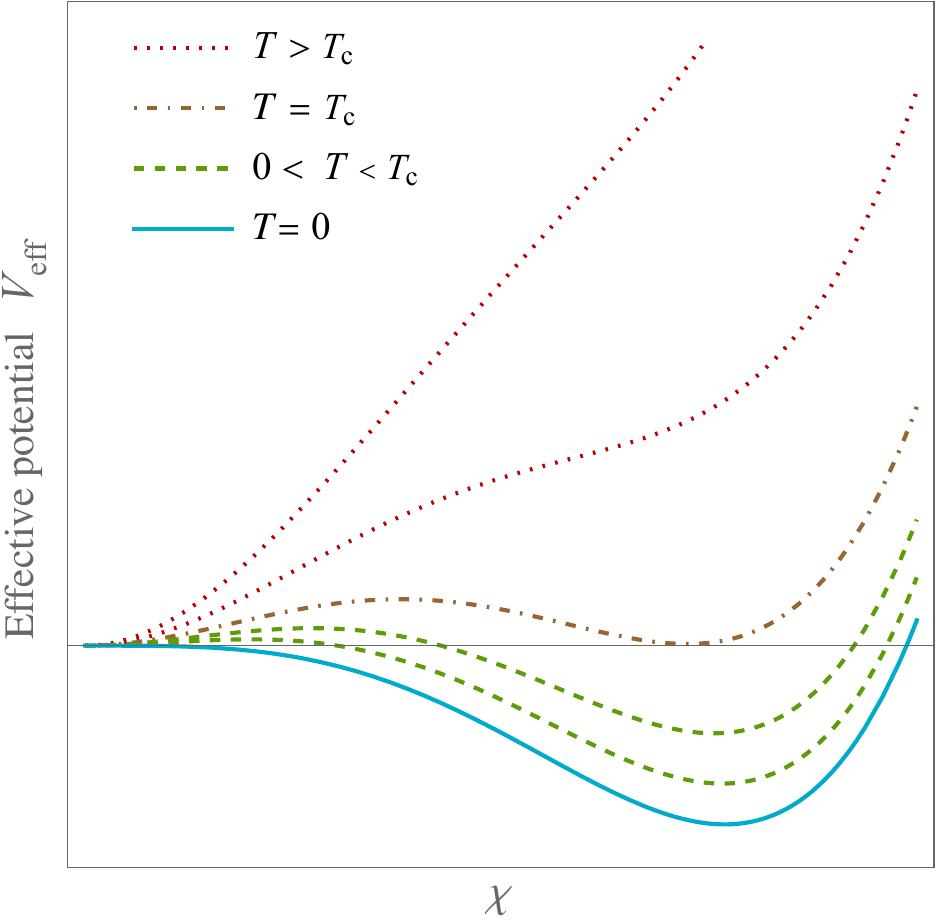}   
    \caption{\em A qualitative picture of the  temperature-dependent effective potential  corresponding to a FOPT: the two minima associated with the two phases are separated by a potential barrier; $T_c$ is the critical temperature. Figure reproduced from Ref.~\cite{Salvio:2023qgb}.}\label{CWpotf}
  \end{center}
\end{figure}

 The PTs induced by radiative symmetry breaking are generically first order, namely of the type illustrated in Fig.~\ref{CWpotf}, as we now show. 
Note that the first three derivatives of the quantum part of the effective potential, $V_q(\chi)$ in~(\ref{CWpot}), vanishes at the origin, $\chi=0$. On the other hand,  $J_B(x)$ and $J_F(x)$ feature in their small-$x$ expansion a term linear in $x$ with a coefficient that is positive in $J_B$ and negative in $J_F$ (see Eqs.~(\ref{JBdef}) and~(\ref{JFdef})). Those  coefficients can be easily recovered: 
\bea J_B(x) &=& J_B(0)+J_B^{(1)} x + ...,  \\
J_F(x) &=&J_F(0)+J_F^{(1)} x + ... , \eea
where the dots are the $\mathcal{O}(x^2)$ terms and
\bea J_B^{(1)}&=&\frac12 \int_0^{\infty} dp\frac{p}{e^p-1},  \\
J_F^{(1)} &=&-\frac12 \int_0^{\infty} dp\frac{p}{e^p+1} . 
\eea 
Note that $J_B^{(1)}$ and $J_F^{(1)}$  are obviously positive and negative, respectively. Indeed, computing them explicitly one finds 
\be J_B^{(1)} =\frac{\pi^2}{12}, \qquad J_F^{(1)} = -\frac{\pi^2}{24}. \label{JFB1}\ee 
Since $J_B$ and $J_F$ appear in the effective potential in the way described by (\ref{VeffSumm}), this implies that the effective potential has a minimum at the origin, $\chi=0$, at {\it any} non-vanishing temperature.  In other words thermal corrections render $\chi=0$ at least a local minimum. Going to small enough temperatures the absolute minimum should be approximately the $T=0$ one given by the potential $V_q$, but since there is always a positive quadratic term thanks to the finite-$T$ contributions the full effective potential  always features a barrier between $\chi=0$ and $\chi=\chi_0$. 
So the PT is always of first order. Note that this reasoning does not assume the presence of a cubic term, $\sim \chi^3$, in the effective potential, which emerges when bosonic fields are coupled to $\chi$ (see Eq.~(\ref{JBdef})), because it also holds when only fermion fields interact with $\chi$.

By definition of the critical temperature $T_c$, the absolute minimum of the effective potential is at $\langle \chi\rangle=0$ for $T>T_c$, while, for $T<T_c$, is at a non-vanishing temperature-dependent value. In the latter case the decay rate per unit of volume, $\Gamma$, of the false vacuum $\langle\chi\rangle=0$ into the true vacuum $\langle\chi\rangle\neq 0$ occurs via quantum and thermal tunnelling through the barrier and can be computed with the formalism of~\cite{Coleman:1977py,Callan:1977pt,Linde:1980tt,Linde:1981zj}: 
\be  \Gamma\sim \exp(-S)\, ,\label{Gamma}\ee
where $S$ is the action
\be S=4\pi\int_0^{1/T} dt_E \int_0^\infty dr r^2\left(\frac12 \dot\chi^2+\frac12 \chi'^2+\bar V_{\rm eff}(\chi,T)\right),  \qquad \bar V_{\rm eff}(\chi,T)\equiv V_{\rm eff}(\chi,T)-V_{\rm eff}(0,T),\label{SVeff}\ee
evaluated at the bounce, which is  the solution of the  differential problem~\cite{Salvio:2016mvj,Shoji:2025nvj} 
\bea && \qquad \ddot\chi+\chi''+\frac{2}{r}\chi'= \frac{d\bar V_{\rm eff}}{d\chi}, \label{bounceProbE} \\ \dot\chi(r,0)=0, &&\,\dot\chi(r,\pm 1/(2T))=0, \quad  \chi'(0,t_E)=0, \quad \lim_{r\to \infty}\chi(r,t_E) = 0. \label{bounceProb}\eea
A dot denotes a derivative with respect to the Euclidean time $t_E$ and a prime denotes a derivative with respect to the spatial radius $r\equiv \sqrt{\vec{x}^{\,2}}$. Indeed, the theory at finite $T$ can be formulated as a theory at imaginary time with time period $1/T$ and the boundary conditions in~(\ref{bounceProb}) impose this periodicity. A particular solution of~(\ref{bounceProbE})-(\ref{bounceProb}) is the time-independent bounce, 
\be \chi''+\frac{2}{r}\chi'= \frac{d\bar V_{\rm eff}}{d\chi}, \qquad \chi'(0)=0, \quad \lim_{r\to \infty}\chi(r) = 0, \label{bounceProb3}\ee
for which
\be S =\frac{S_3}{T}, \qquad  \quad S_3 \equiv 4\pi \int_0^\infty dr \, r^2\left(\frac12 \chi'^2+\bar V_{\rm eff}(\chi,T)\right). \label{S3f}\ee
But generically there could also be time-dependent solutions. When the time-independent bounce dominates the decay rate  per unit of volume is~\cite{Linde:1980tt,Linde:1981zj}
\be  \Gamma\approx T^4\exp(-S_3/T). \label{Gamma3}\ee
Also note that $S_3$ evaluated at the time-independent bounce can be simplified through the  arguments of~\cite{Coleman:1977th} to obtain 
\be S_3 =  -8\pi \int_0^\infty dr \, r^{2} \bar V_{\rm eff}(\chi,T). \label{bounceSd}\ee
Using the expression above instead of the one in~(\ref{S3f}) simplifies numerical calculations  because the derivatives of $\chi$ do not appear in  the action.
In general bounce solutions describe bubbles of the true vacuum inside a background of false vacuum.

\subsection{Supercooling}\label{supercool}

 As long as the couplings are small enough to preserve the validity of perturbation theory, in a generic theory with RSB, Eq.~(\ref{eq:Lmatterns}),  when $T$ goes below $T_c$ the scalar field $\chi$ is trapped in the false vacuum $\langle\chi\rangle=0$ until  $T$ is much below $T_c$. In other words the universe features a phase of supercooling~\cite{Witten:1980ez}. To understand why this is always the case, note that if the theory is scale invariant  $\Gamma$ must scale as $T^4$ and, therefore, the smaller $T$,  the smaller $\Gamma$ (being $\Gamma$ non negative). At quantum level, however, scale invariance is broken by perturbative loop corrections, which introduce another dependence on $T$ in $\Gamma$. This dependence, however, is logarithmic and can become sizable only when $T$ is very small compared to the other scale of the problem, $\chi_0$.

We also note that, as a consequence of supercooling, the derivative loop corrections to the effective action can be neglected: at $T=0$ the quantum effective potential is very shallow because it is only due to perturbatively small loop corrections. So higher-derivative corrections are very small. Also two-derivative and potential loop corrections are suppressed by loop factors. At $T\neq0$ supercooling implies that for the relevant temperatures the thermal corrections are very small and so $V_{\rm eff}$ is still  very shallow and derivative corrections are very small.

In order to describe the universe we live in, supercooling should be followed by a period of  reheating. A general analysis in the RSB scenario was performed in~\cite{Rescigno:2025ong}.

 \section{Supercool expansion at leading order}
\label{LO}
 
 If supercooling is strong enough in a generic theory of the form~(\ref{eq:Lmatterns}), to good accuracy\footnote{The accuracy of the approximation will be analyzed in Sec.~\ref{NLO}.}, the full effective action for values of $\chi$ relevant for the PT can be described to good approximation by three and only three parameters: $\chi_0$, $\bar\beta$ and a real and non-negative quantity $g$ defined as 
follows
\be g^2 \chi^2\equiv \sum_b n_bm_b^2(\chi)+\sum_f m^2_f(\chi). 
\label{M2g2def}\ee
In other words $g^2\chi^2$ is the sum of all bosonic squared masses plus the sum of all Weyl-spinor squared masses\footnote{The quantity $g^2 \chi^2$ defined in~(\ref{M2g2def}) does not coincide with the supertrace of the squared-mass matrix because the fermion masses contribute positively  in~(\ref{M2g2def}).}.
All $m_b^2$ and $m_f^2$ are real, non-negative  and proportional to $\chi^2$, so $g^2$ is real, non-negative and $\chi$ independent.
Note that $g$ plays the role of a tree-level ``collective coupling" of $\chi$ with all fields of the theory.

  \subsection{Tunneling}

To show the property mentioned above one can look in detail at the tunneling process. First note that the dominant contributions to the bounce action $S$ are those from field values around the barrier.
 Therefore, we first need to estimate the barrier size, which we can be defined as the field value $\chi_b$ at which $V_{\rm eff}$ equals its value at the false vacuum $\chi=0$:
\be \bar V_{\rm eff}(\chi_b,T)  = 0, \ee 
where $\bar V_{\rm eff}$ is defined in Eq.~(\ref{SVeff}).
Since $\bar V_{\rm eff}$ depends on $T$, the field value $\chi_b$ will be a function of $T$. Now let us write the logarithmic term  in the quantum contribution $V_q$ to $V_{\rm eff}$, Eqs.~(\ref{CWpot}) and~(\ref{VeffSumm}), as follows
\be \log\frac{\chi_b}{\chi_0} -\frac14= \log\frac{\chi_b}{T}-\frac14+\log\frac{T}{\chi_0}. \label{logSplit}\ee 
In the presence of supercooling, $T\ll\chi_0$,
 we expect that neglecting the first two terms in the right-hand-side of Eq.~(\ref{logSplit}) is a good approximation because, unlike $\chi_0$, the field value $\chi_b$ is clearly small when $T\ll \chi_0$,
\be \log\frac{\chi_b}{\chi_0} -\frac14\approx \log\frac{T}{\chi_0}. \label{logApp}\ee 
If so, using the expression of the effective potential in~(\ref{CWpot}) and~(\ref{VeffSumm}), one finds 
\be \frac{\chi_b^4}{T^4} \approx \frac2{\pi^2}\frac{J_T(\chi_b^2/T^2)-J_T(0)}{\bar\beta\log\frac{\chi_0}{T}},  \label{chibT}\ee
where 
\be J_T(\chi^2/T^2) \equiv \sum_b n_b J_B(m_b^2(\chi)/T^2)-2\sum_f J_F(m_f^2(\chi)/T^2).\ee
Eq.~(\ref{chibT}) tells us that supercooling, $T\ll\chi_0$, suppresses the ratio $\chi_b(T)/T$ logarithmically:
\be \frac{\chi_b^4}{T^4} \approx \frac2{\pi^2}\frac{J_T(\chi_b^2/T^2)-J_T(0)}{\bar\beta\log\frac{\chi_0}{T}} \approx \frac2{\pi^2}\frac{J_T'(0)}{\bar\beta\log\frac{\chi_0}{T}}  \frac{\chi_b^2}{T^2} = \frac{g^2}{6\bar\beta \log\frac{\chi_0}{T}}\frac{\chi_b^2}{T^2} \implies \frac{\chi_b^2}{T^2}  \approx \frac{g^2}{6\bar\beta \log\frac{\chi_0}{T}}, \label{chibTorder} \ee
where in the third step Eqs.~(\ref{JFB1}) and~(\ref{M2g2def}) have been used.
 So the approximation in~(\ref{logApp}) is indeed valid.
Looking now at the effective potential in~(\ref{VeffSumm}) we see that if the quantity $\epsilon$ defined by
\be \boxed{\epsilon\equiv  \frac{g^4}{6\bar\beta \log\frac{\chi_0}{T}}}
 \label{CondConv}\ee
is {\it small} 
 we can approximate
\bea J_B(x) &\approx& J_B(0)+\frac{\pi^2}{12} x ,\label{JBapp}  \\
J_F(x) &\approx&J_F(0)-\frac{\pi^2}{24}x, \label{JFapp}\eea
where~(\ref{JBdef}) and~(\ref{JFdef}) have been used. We will refer to this approximation as the supercool expansion at leading order (LO) as higher orders in the $x$ expansions in~(\ref{JBdef}) and~(\ref{JFdef}) are neglected in this approximation; Sec.~\ref{NLO} will illustrate how one can go beyond the LO.  In~(\ref{CondConv}) an extra factor $g^2$ has been conservatively added compared to~(\ref{chibTorder}) because $\chi^2/T^2$ appear in the thermal functions multiplied by some coupling constant.
Note that the approximations in~(\ref{JBapp}) and~(\ref{JFapp}) are not valid for all values of $\chi$, for example, they  are not valid at  $\chi=\chi_0$. But, as we have just shown, they are valid for the field values that are important in the bounce action if a large-enough supercooling occurs ($\epsilon$ small). This is because in this case $g^2 \chi_b^2/T^2$ is small (see the last equation in~(\ref{chibTorder})).
Now, using the approximations in~(\ref{logApp}),~(\ref{JBapp}) and~(\ref{JFapp}), the bounce action can be computed with the effective potential given by
\be \boxed{\bar V_{\rm eff}(\chi,T) \approx \frac{m^2(T)}{2} \chi^2-\frac{\lambda(T)}{4} \chi^4,} \label{barVapp}\ee
where $m$ and $\lambda$ are real and positive functions of $T$ defined by
\be m^2(T) \equiv \frac{g^2 T^2}{12}, \qquad \lambda(T) \equiv \bar\beta \log\frac{\chi_0}{T} \label{mlambdaDef}\ee
and $g^2$ is the collective coupling  defined in~(\ref{M2g2def}). In  the supercool expansion at LO $\bar V_{\rm eff}$ is given by~(\ref{barVapp}).


We can now see that the tunneling process is dominated by the time-independent bounce, which satisfies~(\ref{bounceProb3}). The expression of $\bar V_{\rm eff}$ in~(\ref{barVapp}), together with the form of the bounce problem in~(\ref{bounceProbE})-(\ref{bounceProb}), tells us that the characteristic bounce size $R_b$ is of order $R_b\sim 1/m(T)\gtrsim1/T$, where in the second step we have used the perturbative requirement that $g$ is not too large. This result tells us that  the bounce solutions  are approximately time-independent (see~\cite{Linde:1980tt,Linde:1981zj}). Moreover, for a time-dependent bounce the Euclidean action $S$ turns out to be larger than the one  of the time-independent bounce for all values of $\lambda$ and $g$ (at least in  perturbative setups)~\cite{Salvio:2016mvj}.  So, as anticipated, the tunneling process is dominated by the time-independent bounce.

\begin{figure}[t]
\begin{center}\includegraphics[scale=0.47]{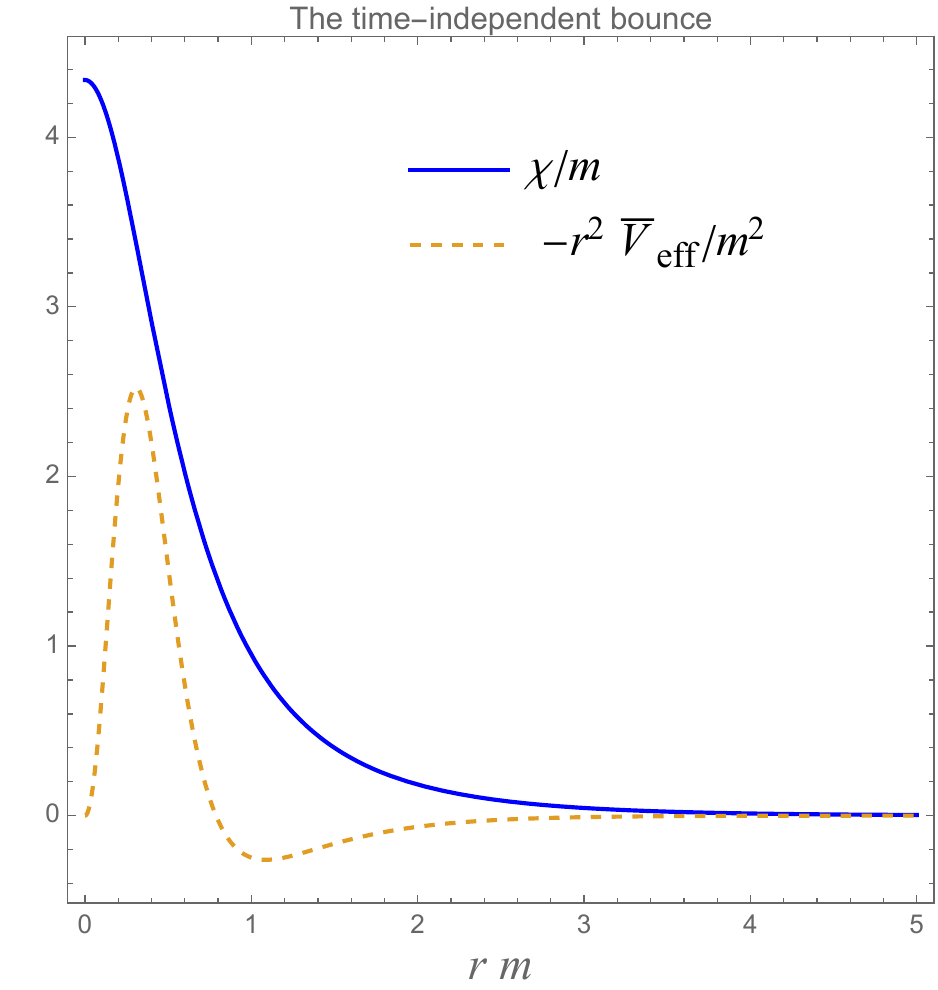}  
    \caption{\em The time-independent bounce and the corresponding integrand function (divided by $8\pi$) appearing in the bounce action, Eq.~(\ref{bounceSd}), for the effective potential $\overline V_{\rm eff}(\chi) = \frac{m^2}{2} \chi^2-\frac{\lambda}{4}\chi^4$ of the supercool expansion at LO and setting $\lambda=1$. Figure reproduced from Ref.~\cite{Salvio:2023qgb}.}\label{BounceSE}
  \end{center}
\end{figure}

Note now that the bounce action $S_3$ computed with the effective potential in~(\ref{barVapp}) is a function of $m$ and $\lambda$ only,  $S_3=S_3(m,\lambda)$. If we rescale $\chi\to\chi/\sqrt{\lambda}$  one finds $S_3(m,\lambda)=S_3(m,1)/\lambda$. Also, using dimensional analysis 
\be S_3=c_3 \frac{m}{\lambda}, \label{S3c3}\ee where $c_3$ is a dimensionless number: computing explicitly the bounce for $\lambda=1$ (see Fig.~\ref{BounceSE}) and its action through Eq.~(\ref{bounceSd}) one finds 
\be c_3=-\frac{8\pi}{m}
 \int_0^\infty dr \, r^2 \left( \frac{m^2}{2} \chi^2-\frac{1}{4} \chi^4\right)
=18.8973... \label{c3value}\ee
 (see also~\cite{Brezin-Parisi,Arnold:1991cv,Salvio:2023qgb} for previous calculations). This quite large value of $c_3$ is due to the geometrical factor of $4\pi$ overall.  Fig.~\ref{BounceSE} shows that for the values of $r$ that give the largest contribution to the bounce action the quartic term is significantly bigger than the quadratic one. 

\subsection{Nucleation temperature}\label{Tn}

During supercooling the energy density  is dominated by the vacuum energy of $\chi$. Therefore, the spacetime is well described by the de Sitter metric and the universe expands exponentially with Hubble rate 
\be H_I = \frac{\sqrt{\bar\beta} \chi_0^2}{4\sqrt{3}\bp},\label{HI}\ee
leading to another inflationary period.
 The bubbles created are   diluted by the expansion of the universe and they cannot collide until $T$ reaches the nucleation temperature $T_n$, which by definition is such that  $\Gamma/H_I^4 \sim 1$ or, equivalently, using the fact that the decay is dominated by the time-independent bounce,
\be  \frac{S_3}{T_n} \approx 4 \log \left(\frac{T_n}{H_I}\right),\label{TnEq0}\ee
By using the expression of $S_3$ in~(\ref{S3c3}) and the definitions in~(\ref{mlambdaDef}) one finds 
\be \frac{c_3 g}{\sqrt{12}\bar\beta\log\frac{\chi_0}{T_n}} \approx 4\log\left(\frac{4\sqrt{3} \bp T_n}{\sqrt{\bar\beta}\,\chi_0^2}\right)\ee
or, equivalently, the approximate quadratic equation
\be   c X - 4 X^2 - a \approx  0\label{TnEq}\ee 
in the variable
\be X\equiv \log\frac{\chi_0}{T_n} \label{Xdef}\ee
and
\be a\equiv \frac{c_3g}{\sqrt{12}\bar\beta}, \quad c\equiv 4\log\frac{4\sqrt{3}\bp}{\sqrt{\bar\beta}\,\chi_0}. \label{caDef}\ee
Recall that $g$ is never negative, the number $c_3$ has the positive value in~(\ref{c3value}) and $\bar\beta>0$, see~(\ref{eq:CWgen}), so $a$ is never negative.

Here we are interested in the solution of Eq.~(\ref{TnEq}) with the smaller $X$, which corresponds to $\Gamma$ reaching $H_I^4$ from below. This solution  gives
\be T_n\approx \chi_0\exp\left(\frac{\sqrt{c^2-16a}-c}8\right). \label{appTn}\ee
Note that in the decoupling limit ($g\to0$, $\bar\beta/g\to0$ and $\chi_0$ fixed) $\sqrt{a}\to\infty$ faster than $c\sim \log a$, the square root in~(\ref{appTn}) develops an imaginary part 
and there is no solution for $T_n$. This happens because in this limit $\Gamma\to 0$ and so can never be of order $H_I^4$. Thus, the existence of a solution of Eq.~(\ref{TnEq}), which determines $T_n$, requires a minimum value of the collective coupling $g$, corresponding to the bound $c^2\geq 16a$; 
however, the smaller $g$ (at $\bar\beta/g^4$ fixed) the larger $\log(\chi_0/T_n)$  and the universe supercools more, at least for realistic and perturbative values of the parameters.
Moreover, in general supercooling is also enhanced by increasing $\chi_0$ at $g$ and $\bar\beta$ fixed, although this effect is very mild as $c$ depends on $\chi_0$ only logarithmically.
In Sec.~\ref{NLO} we will discuss the accuracy of the LO approximations performed in this section and explain how to improve them.

\subsection{Role of gravity in the vacuum decay}\label{RoleGravity}

The gravitational corrections to the false vacuum decay are amply negligible if the symmetry-breaking scale $\chi_0$ is small compared to the Planck mass $M_P$, which is the most interesting case from the  phenomenological point of view. This is because, as we have seen, the temperature and  the typical scales of the bounce are always much below $\chi_0$ and so the gravitational corrections are suppressed by factors much smaller than 
 $\chi_0^2/\bp^2$~\cite{Salvio:2016mvj}, where $\bp$ is the reduced Planck mass defined in terms of the Planck mass $M_P$ by $\bp \equiv M_P/\sqrt{8\pi}$.
 
Moreover, one may wonder whether the effect of the spacetime curvature due to $H_I\neq 0$ can alter the decay rate. In standard Einstein gravity, this may happen if $T_n$ is small enough to be comparable with $H_I$. Whenever a solution for $T_n$ exists, this is typically not the case for realistic and perturbative values of the parameters. On the other hand, if a solution for $T_n$ does not exist  the effect of the spacetime curvature, and quantum fluctuations, can eventually become important in the decay rate~\cite{Kearney:2015vba,Joti:2017fwe,Markkanen:2018pdo,DelleRose:2019pgi}. Furthermore, in modified gravity theories the story can change, because one can introduce non-minimal couplings between $\chi$ and gravity~\cite{Salvio:2023blb}.
 
 \subsection{Strength and inverse duration of the phase transition}

Generically, the strength of the PT is measured by the parameter $\alpha$ defined as the ratio between 
\be \rho(T_n) \equiv \left[\frac{T}{4} \frac{d}{dT} \bar V_{\rm eff}(\langle\chi\rangle,T)-\bar V_{\rm eff}(\langle\chi\rangle,T)\right]_{T=T_n} \label{rhoDef} \ee
and the energy density of the thermal plasma (see~\cite{Caprini:2019egz,Ellis:2019oqb} for more details). In~(\ref{rhoDef}) $\langle\chi\rangle$ corresponds to the absolute minimum of the effective potential, which at $T=T_n$ is not zero.  By definition $\alpha$ is
\be \alpha \equiv \frac{30 \rho(T_n)}{\pi^2 g_*(T_n)T_n^4}, \ee
with $g_*(T)$ being the effective number of relativistic species at temperature $T$.
In the presence of supercooling we then typically have 
\be \rho(T_n) \approx \left[-\bar V_{\rm eff}(\langle\chi\rangle,T)\right]_{T=T_n}  \ee
and
\be \alpha \gg 1,\ee
because the energy density is dominated by the vacuum energy density of $\chi$, as we have seen.
 So in the RSB scenario one has  a very strong PT.

 Another important parameter  is the inverse duration $\beta$ of the PT that, in models with supercooling, is given by~\cite{Caprini:2015zlo,Caprini:2018mtu,vonHarling:2019gme} 
\begin{equation}
	\beta \equiv \bigg[\frac{1}{\Gamma(t)}\frac{d\Gamma(t)}{dt}\bigg]_{t=t_n}, \label{betaDef}
\end{equation}
  where   $t_n$ is the value of the time $t$ when $T=T_n$. Using then $dt = -dT/(TH(T)) $, where $H(T)$ is the Hubble rate at temperature $T$, 
  and the fact 
that the tunneling process is dominated by the time-independent bounce,
\be \beta \approx H_n\left[T\frac{d}{dT}(S_3/T)-4\right]_{T=T_n}, \label{betaH1}\ee 
where $H_n\approx H_I$ is the Hubble rate  at $T=T_n$.

 Now, if supercooling is large enough that  the expression of $S_3$ in~(\ref{S3c3}) is valid,
 one obtains 
\be \frac{\beta}{H_n} \approx \frac{a}{\log^2(\chi_0/T_n)} -4 , \label{betaH2} \ee
where $a$ is defined in terms of $\bar\beta$ and $g$ in~(\ref{caDef}).
  Eq.~(\ref{betaH2}) explicitly shows that the PT lasts more when $T_n/\chi_0$ is smaller. 
  Inserting the expression of $T_n$ given in~(\ref{appTn}) into~(\ref{betaH2}),
  one obtains $\beta/H_n$ in terms  of $\chi_0$, $\bar\beta$  and $g$.

\subsection{Time dependence of the vacuum decay rate}\label{Time dependence of the false-vacuum decay rate}

The decay rate per unit of volume, $\Gamma$, can be expanded for a generic time $t$ around the nucleation time $t_n$, when  the PT starts to be effective. By definition of $T_n$ we have  $\Gamma(t_n)=H(t_n)^4$, where $H$ is the Hubble rate, and the above-mentioned expansion reads 
\begin{equation}
	\label{eq:GammaGen}
    \Gamma(t) = \Gamma(t_n)\text{exp}(\beta(t-t_n)+ \beta_2(t-t_n)^2+ \beta_3(t-t_n)^3+...).
\end{equation}
The parameter $\beta$ has been defined in~(\ref{betaDef}).
Neglecting the coefficients ($\beta_2, \beta_3, ...$) of the higher powers of $t-t_n$  leads to
 \begin{equation}
	\label{eq:Gamma}
    \Gamma(t) \approx H_n^4 e^{\beta(t-t_n)}.
\end{equation}
where $ H_n\equiv H(t_n)$. This approximation is widely used in the context of primordial-black-hole production from FOPTs (see e.g.~\cite{Liu:2021svg,Gouttenoire:2023naa}). We now discuss its accuracy.

To this purpose let us consider the phase of the evolution of the universe since a time close to $t_n$. This is because vacuum decay is not effective at much earlier times. We can approximate $H(t)\approx H_I$ (constant in time) since the spacetime exponentially expand at those times before vacuum decay occurs: here we use supercooling, which tells us that the energy density is dominated by the vacuum contribution at $t_n$. 
Using also $dt = -dT/(TH) \approx -dT/(TH_I)$ we find
\be T(t) \approx T_n e^{-H_I(t-t_n)}. \label{tTcon} \ee
In the approximation of~(\ref{S3c3})
\be \frac{S_3}{T}\approx c_3 \frac{m}{T\lambda} = \frac{c_3 g}{\sqrt{12}\bar\beta \log\frac{\chi_0}{T}} \equiv \frac{a}{X+\log\frac{T_n}{T}}\approx \frac{a}{X+H_I(t-t_n)},\ee
where  we used~(\ref{tTcon}) in the last step and  $X$ is defined in~(\ref{Xdef}).
Now,  inserting this result and~(\ref{tTcon}) in~(\ref{Gamma3}) gives
\be  \Gamma(t) \approx T_n^4\exp\left(-\frac{a}{X+H_I(t-t_n)}-4 H_I(t-t_n)\right) \ee
and expanding for $t$ around $t_n$ the first term in the  exponent, 
\be \frac{1}{1+\frac{H_I(t-t_n)}{X}} =  1 -\frac{H_I(t-t_n)}{X}+\left(\frac{H_I(t-t_n)}{X}\right)^2+... + (-1)^k\left(\frac{H_I(t-t_n)}{X}\right)^k + ...  \label{TayE}\ee 
we can clearly see that the constant and ${\cal O}(t-t_n)$ terms reproduce the right-hand side of~(\ref{eq:Gamma}) because $a/X$ equals $S_3/T$ at $t=t_n$ and 
\be \frac{a H_I}{X^2} - 4H_I \approx \beta \ee
according to~(\ref{betaH2}). The corrections are small as long as 
\be H_I(t-t_n) \ll X, \label{GappVal}\ee 
Now, in the approximations used, $X$ is rather large. Indeed, $X$ is typically more than an order of magnitude larger than $1/\epsilon$ evaluated at $T_n$ (as $\bar\beta$ is typically of order $g^4/(4\pi)^2$). 
So, the condition in~(\ref{GappVal}) can remain valid for a significantly large number of e-folds, $H_I(t-t_n)$, and in that period~(\ref{eq:Gamma}) is a good approximation. This approximation  becomes increasingly accurate as the supercooling (namely $X$) increases.

\section{Supercool expansion beyond the leading order}\label{NLO}

The approximation performed in~(\ref{JBapp})-(\ref{JFapp}) generically corresponds to neglecting terms of order $\sqrt{\epsilon}$ (where $\epsilon$ is defined in~(\ref{CondConv})). Since we eventually need to set $T=T_n$, see Eq.~(\ref{betaH1}), the approximation in~(\ref{logApp}), on the other hand,  corresponds to neglecting terms of relative order $(\log X)/X$, which is smaller than $\sqrt{\epsilon}$ because, since $\epsilon$ is small,  $X\gtrsim g^4/(6\bar\beta) = \epsilon X$, which is large because $\bar\beta$ is loop suppressed\footnote{The approximation in~(\ref{logApp}) consists in neglecting terms of relative order $1/X$ and $\log(\epsilon/g^2)/X$. The order of magnitude of the former is obviously not larger  than $(\log X)/X$, but the same is true for the  latter: apparently it might become larger than $(\log X)/X$ when $g\to 0$, but, as we have seen around~(\ref{appTn}), in this limit there is no solution for $T_n$ and, in any case, $X$ asymptotically goes as an inverse power of $g$ because $\bar\beta \sim g^4/(4\pi)^2$. }. 
 What we are doing here is a small-$\epsilon$ expansion (a ``supercool expansion") and what we have done so far is the LO approximation, namely so far we  have worked modulo terms of relative order $\sqrt{\epsilon}$.
In this section we discuss how to improve the LO result by calculating higher-order corrections in the supercool expansion. We will focus on the properties that are significantly changed going beyond the LO. 

The first step is to calculate the corrections of relative order $\sqrt{\epsilon}$ with respect to the LO approximation, which
 corresponds to working at next-to-leading order (NLO)  in the supercool expansion. This can be done by improving the approximation in~(\ref{JBapp})-(\ref{JFapp}): including the term of order\footnote{This term was overlooked in Ref.~\cite{Witten:1980ez} (see Eq.~(3) there).} $x^{3/2}$ in the expansion of $J_B(x)$ in~(\ref{JBdef}), 
\bea J_B(x) &\approx& J_B(0)+\frac{\pi^2}{12} x-\frac\pi{6} x^{3/2} ,\label{JBappnlo}  \\
J_F(x) &\approx&J_F(0)-\frac{\pi^2}{24}x. \label{JFappnlo}\eea
Since for field values that are relevant for the bounce action $x$ is at most of order $\epsilon$ (see Eqs.~(\ref{chibTorder}) and~(\ref{VeffSumm})), this corresponds to neglecting terms of order not larger than $\epsilon\log \epsilon$ relatively to the LO. So the approximation in~(\ref{logApp}) and dropping terms of order $X\log X$  in the equation for the nucleation temperature are still allowed even at NLO.

The effective potential at NLO, therefore, includes a cubic term in the field $\chi$ and reads
\be \boxed{\bar V_{\rm eff}(\chi,T) \approx \frac{m^2(T)}{2} \chi^2-\frac{k(T)}{3}\chi^3-\frac{\lambda(T)}{4} \chi^4,} \label{barVnlo}\ee
where $m^2$ and $\lambda$ are defined 
in~(\ref{mlambdaDef}), 
\be k(T)\equiv \frac{\tilde g^3 T}{4\pi}, \label{kdef}\ee
and $\tilde g$ is a non-negative real parameter  defined as follows:
\be \tilde g^3\chi^3 \equiv \sum_b n_bm_b^3(\chi). \label{gtdef}\ee
Here the $m_b^3$ are the cube of the background-dependent bosonic masses, which are all real, non-negative  and proportional to $\chi$. So $\tilde g^3$ is real, non-negative and independent of $\chi$; it is an extra parameter that is needed to describe this scenario in a model-independent way at NLO. In general we have\footnote{This is because, given two positive numbers $p$ and $q$ with $p<q$, the $p$-norm of a vector in $\mathbb{R}^n$ is never smaller than its $q$-norm.}
\be \tilde g\leq g. \label{disggt}\ee

 In order to understand why the term cubic in $\chi$ in~(\ref{barVnlo}) can be considered as a small correction in the supercool expansion, one can rescale $\chi\to \chi/\sqrt{\lambda}$ in the bounce action, Eq.~(\ref{SVeff}), to find 
\be S = \frac{4\pi}{\lambda}\int_0^{1/T} dt_E \left[ \int_0^\infty dr \, r^2\left(\frac12 \dot\chi^2+\frac12 \chi'^2+\frac{m^2}{2} \chi^2-\frac{1}{4} \chi^4\right) -\frac{k}{3\sqrt{\lambda}} \int_0^\infty dr \, r^2\chi^3 \right].\label{SNLO}\ee 
Since $\lambda=\bar\beta\log(\chi_0/T)$ and eventually we need to set $T=T_n$, we explicitly see that the term proportional to $k$ has relative order at most $\sqrt{\epsilon}$ times a number smaller than one, $\approx 1/(\sqrt{2}\pi)$ (where the LO result $S_3\approx 4\pi gT/(\sqrt{12}\lambda)$ and~(\ref{disggt}) have been used). This small number helps the convergence of the supercool expansion. Working at NLO, we can substitute $\chi$ with the solution of the  LO bounce problem both in the second integral in the square bracket (because suppressed by $\sqrt{\epsilon}$) and in the first one (because the first variation of the action around a solution of the field equations vanishes). So one finds
\be S_3 = \frac{1}{\lambda}\left(c_3 m -\tilde c_3\frac{k}{3\sqrt{\lambda}}\right), \label{S3k0}\ee
where $c_3$ is given in~(\ref{c3value}),
\be \tilde c_3 \equiv 4\pi \int_0^\infty  dr \,  r^2\chi_{\rm LO}^3 \ee
and $\chi_{\rm LO}$ is the LO bounce configuration. A numerical calculation gives $\tilde c_3= 31.6915 ... \,$. Again, like in~(\ref{c3value}), we find a rather large value because of the  geometrical overall factor of $4\pi$.

Having obtained the NLO bounce action we can now improve the LO equation in~(\ref{TnEq}), which gives $T_n$. Using~(\ref{TnEq0}), one obtains
\be   c X - 4 X^2 - a +\frac{\delta}{\sqrt{X}} \approx 0,\label{TnEqNLO0}\ee
where $a$ and $c$ are defined in~(\ref{caDef}) and
\be \delta \equiv \frac{\tilde c_3 \tilde g^3}{12\pi \bar\beta^{3/2}}. \ee
To find a solution for $T_n$ one can again proceed perturbatively: let  $X_{\rm LO}$ be the LO solution, namely
\be X_{\rm LO} = \frac{c-\sqrt{c^2-16a}}8,\ee 
the NLO solution can be obtained by substituting $X$ with $X_{\rm LO}$ only in the part proportional to $\delta$ in Eq.~(\ref{TnEqNLO0}) and then solving with respect to $X$. This leads to the following NLO expression for $T_n$:
\be T_n\approx \chi_0\exp\left(\frac{\sqrt{c^2-16(a-\delta/\sqrt{X_{\rm LO}})}-c}8\right). \label{appTnnlo}
\ee

Let us now determine the NLO expression for $\beta/H_n$, given by~(\ref{betaH1}). To this purpose one can use
\be \frac{S_3}{T} = \frac{a}{\log\frac{\chi_0}{T}} \left(1-\frac{\delta/a}{\sqrt{\log\frac{\chi_0}{T}}}\right) \ee 
to compute~(\ref{betaH1}). Dropping terms of relative order smaller than $\epsilon$ that are negligible even at NLO, 
\be \frac{\beta}{H_n} \approx \frac{a}{\log^2(\chi_0/T_n)}\left(1-\frac{3\delta/a}{2\sqrt{\log(\chi_0/T_n)}}\right) -4. \label{betaH2nlo} \ee
Interestingly, the NLO correction reduces $\beta$ and so renders the spectrum of GWs larger~\cite{Salvio:2023qgb}.

One can then go ahead with the supercool expansion. The next step, the next-to-next-to-leading order (NNLO), would consist in including terms of order $\epsilon \log\epsilon$ and $\epsilon$. This would require including the $x^2\log x$ and $x^2$ terms in the small-$x$ expansion of $J_B(x)$ and $J_F(x)$, given in~(\ref{JBdef})-(\ref{JFdef}), and considering the field-dependent logarithm in~(\ref{logSplit}). Note, however, that  the $\log a_B$ and $\log a_F$ contributions (and thus the $x^2$ terms) in~(\ref{JBdef})-(\ref{JFdef}) can be ignored because they lead to a temperature-independent term proportional to $\chi^4$ in the potential: as a result they can be absorbed in a redefinition of the renormalization-scheme-dependent coefficient $a_s$ in~(\ref{Vqs}).
However, in doing these improvements one  needs to include extra parameters beyond $\chi_0$, $\bar \beta$ $g$ and $\tilde g$. Moreover, going to this higher level of precision would also require estimating all functional determinants appearing in the one-loop prefactor of $\Gamma$ given in~\cite{Linde:1980tt,Linde:1981zj}. So at NNLO one needs to correct Eq.~(\ref{TnEqNLO0})
for the nucleation temperature $T_n$ in a non-trivial way.  Because of the small numbers in front of $x^2\log x$ terms in Eqs.~(\ref{JBdef})-(\ref{JFdef}) 
and the fact that the terms $\log(\chi_b/T)-1/4$ in~(\ref{logSplit}) are at most of order $\log X/X< \epsilon \log\epsilon$ (relatively to the LO)  
one expects that their inclusion would generically lead to very small corrections as long as $\epsilon$ is small. 
A complete analysis of the NNLO and higher orders is not yet present in the literature.

Let us conclude this section with a remark on the discussion in Sec.~\ref{Time dependence of the false-vacuum decay rate} regarding the time dependence of the vacuum decay rate. That 
discussion  is an order-of-magnitude estimate and, as such, it remains valid considering the higher-order corrections in the supercool expansion, as long as, of course, $\epsilon$ remains small. Indeed, the latter condition is the general condition for the validity of the supercool expansion.

\section{Extending the validity of the  supercool expansion}\label{impro}

In this section we study when and how one can extend the validity of the supercool expansion to cases where $\epsilon$ is of order one, 
\be \boxed{ \epsilon \sim 1.} \ee
We will focus on the properties that are significantly changed going from small $\epsilon$ to an order-one $\epsilon$. 

\subsection{Several degrees of freedom}\label{Several degrees of freedom}

The expansion developed in Secs.~\ref{LO} and~\ref{NLO} in general works for $\epsilon$ small. However, it also holds for values of $\epsilon$ of order one if there are several degrees of freedom, say $N$, with dominant couplings (all of the same order of magnitude, say $\tau$) to the flat-direction field $\chi$. Indeed,   in this case $g$ defined in~(\ref{M2g2def}) scales as $g \sim \sqrt{N}\tau$, while $\tilde g$ defined in~(\ref{gtdef}) scales as $\tilde g \lesssim \sqrt[3]{N}\tau$, and so 
$\tilde g^3/g^3 \lesssim1/\sqrt{N}$: the inequality here is due to the fact that $\tilde g$ receives only bosonic contributions, while $g$ receives both bosonic and fermionic contributions. As a result, the extra cubic term in the bounce action of Eq.~(\ref{SNLO}) gets an extra suppression factor (see Eq.~(\ref{kdef})), which is at least as small as $1/\sqrt{N}$. On the other hand, 
\begin{itemize}
\item since $1/X = 6\bar\beta\epsilon/g^4$, for order-one $\epsilon$ the quantity $1/X$ is still small because $\bar\beta$ is loop suppressed and so the approximation in~(\ref{logApp}) is still good,
\item truncating the small-$x$ expansions in~(\ref{JBdef}) and~(\ref{JFdef}) up to the $x^{3/2}$ term is still justified because the higher-order terms involve smaller and smaller coefficients\footnote{One can check that by looking at the full expansions of $J_B(x)$ and $J_F(x)$ provided, for example, in~\cite{Quiros:1994dr}.}, with the coefficient of the $x^2\log x$ term being already quite small ($\sim 1/32$). Indeed, we recall that  the $\log a_B$ and $\log a_F$ contributions in~(\ref{JBdef})-(\ref{JFdef}) can be  absorbed in a redefinition of the renormalization-scheme-dependent coefficient $a_s$ in~(\ref{Vqs}).
\end{itemize}

\subsection{Improved supercool expansion}\label{Improved supercool expansion}

If the number of degrees of freedom with a dominant coupling to $\chi$ is too small, one instead has $\tilde g\approx g$ and, in this case, the expansion of Secs.~\ref{LO} and~\ref{NLO} breaks down for order-one values of $\epsilon$ (although it still holds for small $\epsilon$).

\subsubsection{Bounce solution and action}

 In order to extend the class of theories that can be described by the supercool expansion one should, therefore, include the cubic term in~(\ref{barVnlo}) in the non-perturbative computation of the bounce action. However, the other corrections can still be treated as perturbations; indeed, they are still small as long as $\epsilon$ is at most of order one, as we have  seen in Sec.~\ref{Several degrees of freedom}. This improvement is called the ``improved supercool expansion". Let us explain how to construct it.

The expression of $\bar V_{\rm eff}$ in~(\ref{barVnlo}), together with the form of the bounce problem in~(\ref{bounceProbE})-(\ref{bounceProb}), still tells us that the characteristic bounce size $R_b$ is of order $R_b\sim 1/m(T)\gtrsim1/T$, where in the second estimate we have used the perturbativity condition that $g$ is not too large. Indeed, $R_b$ can be read from the large-$r$ limit of the bounce solution and in this limit the last condition in~(\ref{bounceProb}) tells us that only the quadratic term in~(\ref{barVnlo}) matters. Therefore, the bounce solutions  are approximately time-independent even including the cubic term in~(\ref{barVnlo}) in the non-perturbative computation of the bounce action.

\begin{figure}[t]
\begin{center}
  \includegraphics[scale=0.44]{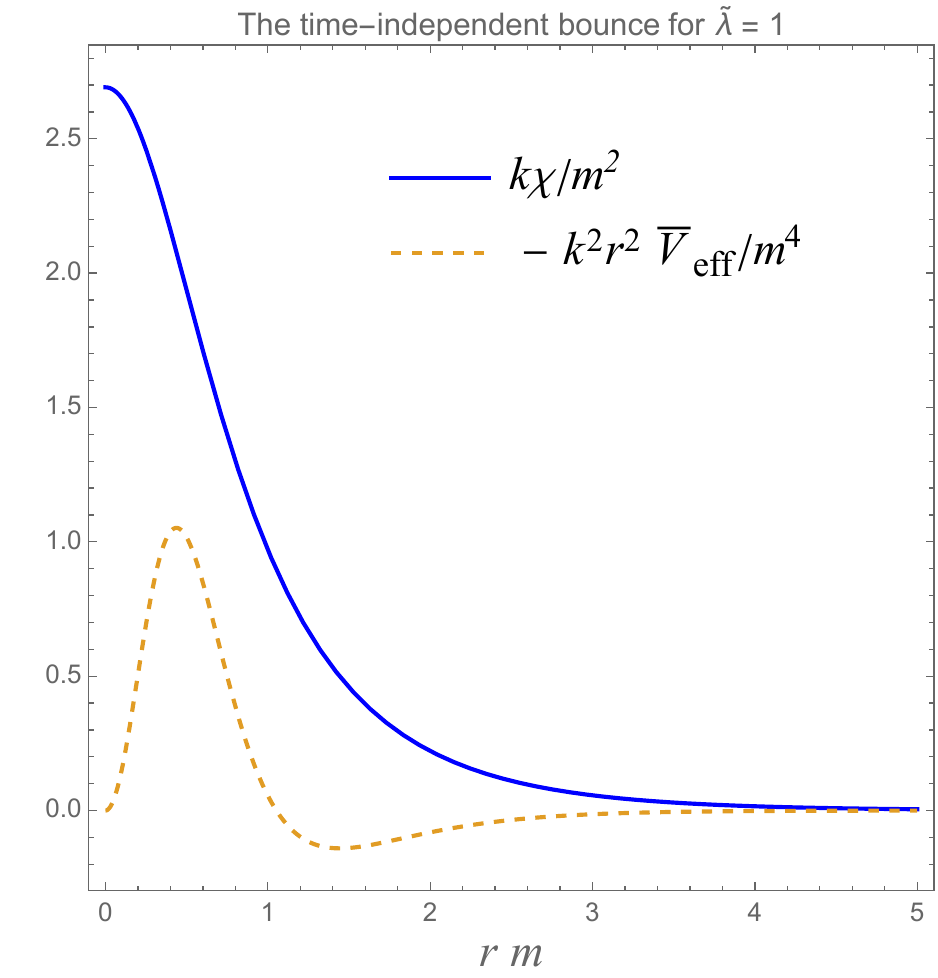}  \hspace{1cm} \includegraphics[scale=0.44]{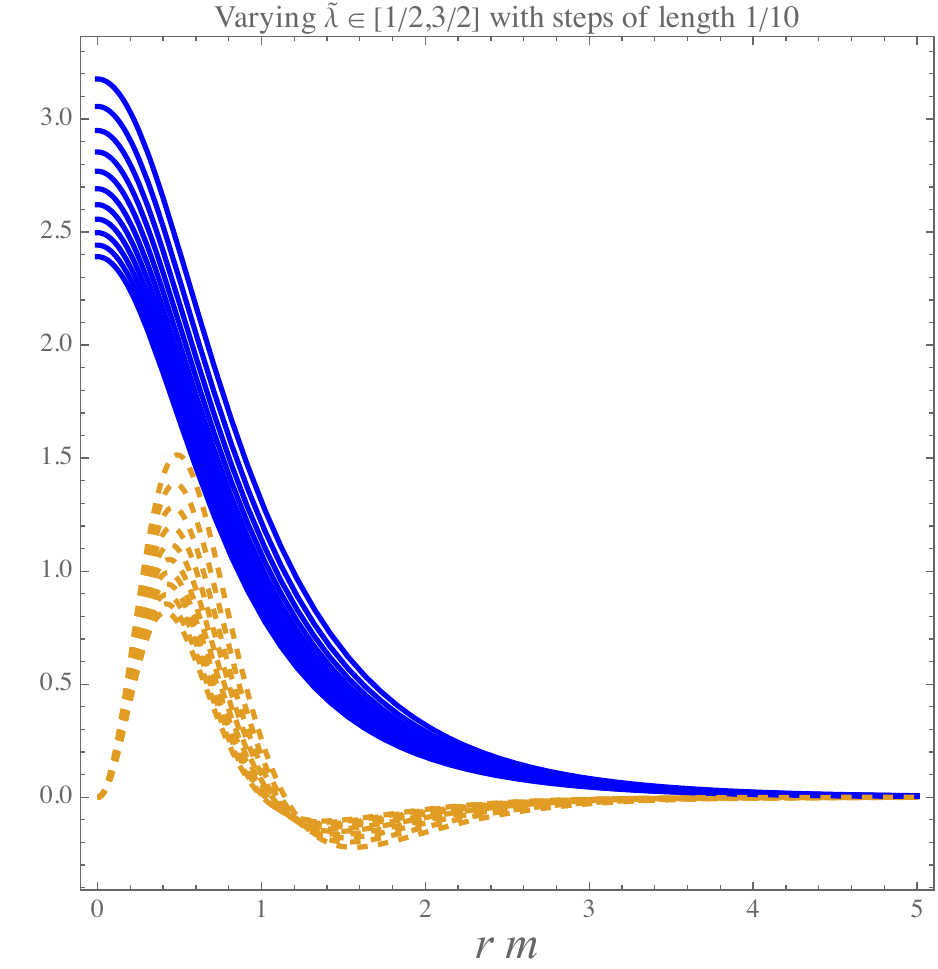}  
    \caption{\em  The relevant bounce and the corresponding integrand function (divided by $8\pi  L \xi^2$) appearing in the bounce action in~(\ref{S3improS}) for the effective potential (\ref{barVnlo}) and varying $\tilde \lambda\equiv  \lambda m^2/k^2$. The maximal height of the curves increases by decreasing $\tilde\lambda$.}\label{kappasetting}
  \end{center}
\end{figure}

Looking then at~(\ref{S3f}) and redefining~\cite{Levi:2022bzt} $r\equiv L\rho$  and $\chi\equiv \xi \varphi$ one obtains the bounce action for the new radial  and field variables $\rho$ and $\varphi$, respectively, 
\be S_3 \equiv 4\pi L\xi^2\int_0^\infty d\rho \, \rho^2\left(\frac12 \left(\frac{d\varphi}{d\rho}\right)^2+\tilde V_{\rm eff}(\varphi,T)\right), \ee 
where
\be \tilde V_{\rm eff}(\varphi,T)\equiv \left(\frac{L}{\xi}\right)^2 \bar V_{\rm eff}(\chi,T).\ee
By evaluating at the bounce solution one then finds, like in~(\ref{bounceSd}), a simplified bounce action
\be S_3 =  -8\pi L \xi^2 \int_0^\infty d\rho \, \rho^{2} \tilde V_{\rm eff}(\varphi,T). \label{S3improS}\ee
Choosing now
\be L = \frac1{m}, \qquad \xi = \frac{m^2}{k}, \ee
with $m$ and $k$ defined in~(\ref{mlambdaDef}) and~(\ref{kdef}), respectively, gives
\bea \tilde V_{\rm eff}(\varphi,T) &=& \frac12 \varphi^2-\frac13 \varphi^3 -\frac{\tilde \lambda}{4}\varphi^4, \\
S_3 &=&-\frac{8\pi m^3}{k^2}\int_0^\infty d\rho \, \rho^{2} \left( \frac12 \varphi^2-\frac13 \varphi^3 -\frac{\tilde \lambda}{4}\varphi^4\right)  \eea
with 
\be \tilde \lambda \equiv  \frac{\lambda m^2}{k^2} <0 \label{deflambdat} \ee 
and $\lambda$ defined in~(\ref{mlambdaDef}).
The quantity $\tilde \lambda$ can also be rewritten by using~(\ref{mlambdaDef}) and~(\ref{kdef}) as follows
\be \tilde \lambda(T) =  \frac{(4\pi)^2 \bar\beta }{12\, \tilde g^6/g^2} \log(\chi_0/T) \ee 
that depends on $T$ only through $\log(\chi_0/T)$.  Using the definition of $\epsilon$ in~(\ref{CondConv}) one obtains  
\be \tilde \lambda = \frac{2\pi^2}{9\epsilon} \frac{g^6}{\tilde g^6}, \label{lambda-eps}\ee
and recalling the bound in~(\ref{disggt})
\be \tilde \lambda \geq \frac{2\pi^2}{9\epsilon}.\ee
So the small-$\epsilon$ expansion of Secs.~\ref{LO} and~\ref{NLO} corresponds to $\tilde \lambda$ large. Here we are interested in setting $\epsilon$ of order one and $\tilde g \approx g$, when that expansion breaks down. Thus we are interested in finite values of $\tilde \lambda$ around $1$. 
In Fig.~\ref{kappasetting} the time-independent bounces for $\tilde \lambda\in[1/2,3/2]$ are shown, together with  $-\rho^{2} \tilde V_{\rm eff}$, which appears in the integrand of the bounce action in~(\ref{S3improS}).

We are not able to compute $S_3$ analytically as a function of $\tilde \lambda$. However, one can compute the bounce and the corresponding $S_3$ for several values of $\tilde \lambda$ and then perform a fit~\cite{Adams:1993zs,Sarid:1998sn,Levi:2022bzt}. Doing so one can check that
\be S_3 =  \frac{27\pi m^3}{2k^2} \frac{1+\exp(-1/\sqrt{\tilde \lambda})}{1+\frac92 \tilde \lambda} =27\pi m^3 \frac{1+\exp(-k/(m\sqrt{\lambda}))}{2k^2+9\lambda m^2} \label{S3k} \ee 
reproduces the numerical calculations at the $\sim 1\%$ level for the values of $\tilde \lambda$ we are interested in. The result in~(\ref{S3k}) was found by~\cite{Levi:2022bzt} in a specific setup.
Its validity has been established in a model-independent way within the improved supercool expansion in~\cite{Salvio:2023ext}.

\subsubsection{Nucleation temperature}\label{Tnimpro}

\begin{figure}[t]
\begin{center}
\hspace{-0.4cm}  
\includegraphics[scale=0.59]{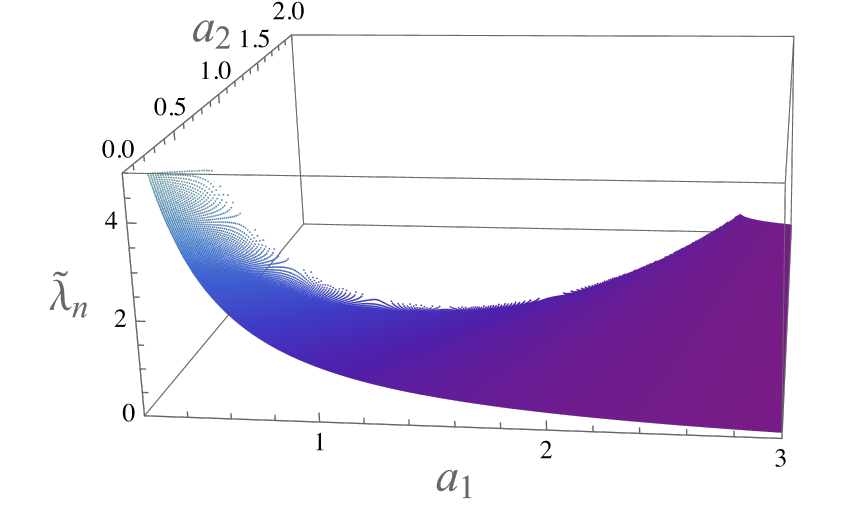}  \hspace{0.5cm}
 \includegraphics[scale=0.36]{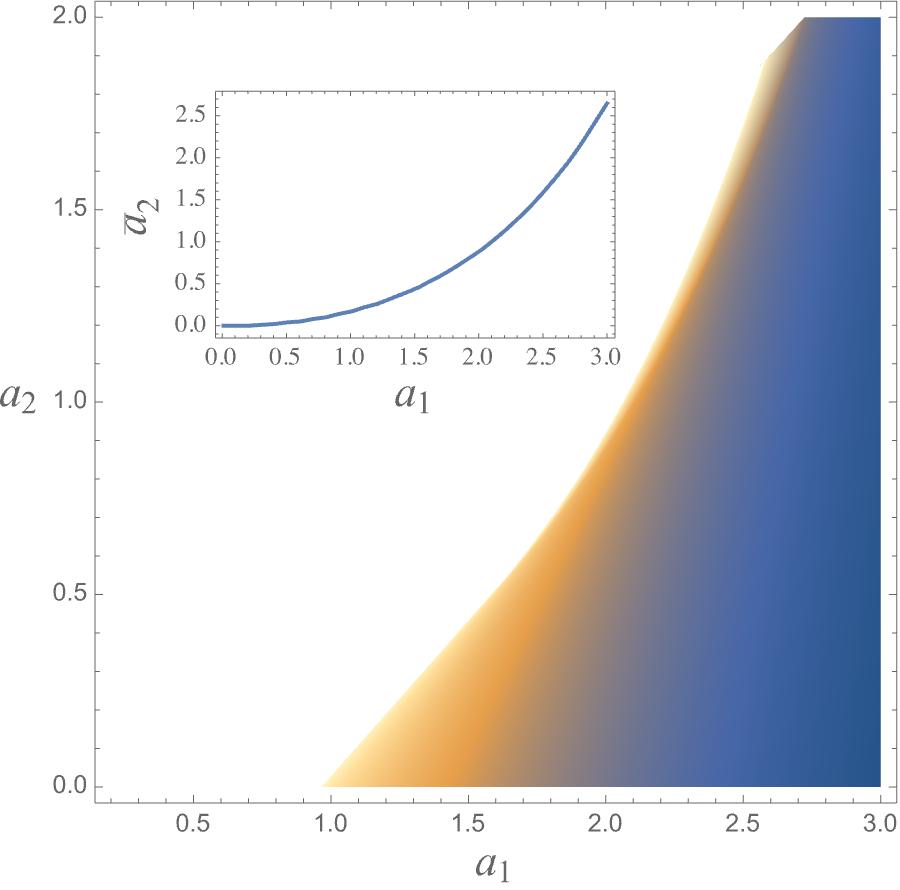}\includegraphics[scale=0.4]{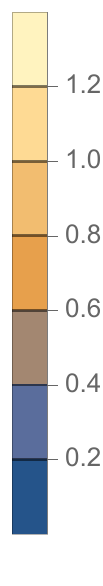}  
    \caption{\em The  solution $\tilde \lambda_n$ of Eq.~(\ref{lteq}) as a function of $a_1$ and $a_2$ defined in~(\ref{a1a2}). The inset in the right plot gives the maximal value of $a_2$ for a given $a_1$ such that the  solution $\tilde \lambda_n$ exists. Using the definitions of $\tilde \lambda$ and $\lambda$ in (\ref{deflambdat}) and (\ref{mlambdaDef}) one can extract the nucleation temperature. Figure reproduced from Ref.~\cite{Salvio:2023ext}.}
    \label{lambdat}
  \end{center}
\end{figure}

Inserting Expression~(\ref{S3k}) in Eq.~(\ref{TnEq0}) for the nucleation temperature $T_n$ leads to a non-polynomial equation in $\tilde \lambda$:
\be a_1-a_2 \tilde \lambda = F(\tilde \lambda) \equiv \frac{1+\exp(-1/\sqrt{\tilde \lambda})}{2/9+\tilde\lambda}, \label{lteq} \ee
where 
\be a_1 \equiv \frac{c\,  c_3 k^2}{3\pi a\, \bar\beta\, m^2}, \qquad a_2 \equiv \frac{4 c_3 k^4}{3\pi a\,  \bar\beta^2m^4}, \label{a1a2}\ee 
$c_3$ is given in~(\ref{c3value}) and $a$ and $c$ are defined in~(\ref{caDef}). Here we are interested in the smallest real and positive solution $\tilde \lambda_n\equiv \tilde\lambda(T_n)$ of Eq.~(\ref{lteq}) for which the straight line $a_1-a_2\tilde \lambda$ reaches $F(\tilde \lambda)$ from below in increasing $\tilde \lambda$ (that corresponds to $\Gamma$ reaching $H_I^4$ from below). 
Such a solution does not always exist for any $a_1$ and $a_2$. First, one needs $a_1\leq F(0) =9/2$; second, for each given $a_1$ the parameter $a_2$ must me smaller than a certain critical value $\bar a_2(a_1)$, which is given in the inset of the right plot of Fig.~\ref{lambdat}.  Fig.~\ref{lambdat} also shows, as a function of $a_1$ and $a_2$, the solution $\tilde\lambda_n$ (when it exists), which has been  obtained numerically.  Tables with the numerical determination of $\bar a_2$ as a function of $a_1$ and of $\tilde \lambda_n$ as a function of $a_1$ and $a_2$ can be found at \cite{dataset}. Once we fix the parameters $g$, $\bar\beta$, $\chi_0$ and $\tilde g$ the quantities $a_1$ and $a_2$ as well as $\tilde \lambda_n$ and thus the nucleation temperature $T_n$ are fixed.

Using the obtained solution $\tilde \lambda_n$ one can check that the PT strength parameter $\alpha$ is large in an RSB PT  for realistic and perturbative values of the parameters, even in the improved supercool expansion. Thus, the plasma effects (such as those studied in Refs.~\cite{Bodeker:2009qy,Bodeker:2017cim}) can be neglected in this particular scenario.

Let us now make a remark regarding the time dependence of the false-vacuum decay rate. In Sec.~\ref{Time dependence of the false-vacuum decay rate} an estimate of such time dependence was provided in the supercool expansion. Since that was an order-of-magnitude estimate, it remains valid here: 
indeed, in the improved supercool expansion, since $\epsilon$ is at most of order 1 and $\bar\beta$ is typically of order $g^4/(4\pi)^2$, the quantity $X$ is still rather large, typically of order of several tens. This has also been numerically checked with the fit in~(\ref{S3k}). 
This model-independent result agrees with a computation performed in  specific RSB models in the context of PBH production (see e.g.~\cite{Gouttenoire:2023pxh}).

 \subsubsection{Duration of the phase transition}\label{Duration of the phase transition}

 Using the expression of $S_3$ in~(\ref{S3k}), we obtain a formula for the inverse duration of the PT:
 \be \frac{\beta}{H_n} \approx \frac{\pi^3g^5}{6\sqrt{3}\tilde g^8}  \frac{(4\pi)^2\bar\beta}{\tilde g^4}(-F'(\tilde \lambda_n)) -4, \label{betaOHimpro}\ee
 where $F'$ is the derivative of $F$ defined in~(\ref{lteq}) with respect to $\tilde \lambda$; note that $F$ is a monotonic decreasing function of $\tilde \lambda$ so the quantity $-F'$ that appears in~(\ref{betaOHimpro}) is positive. Numerical calculations of $\beta/H_n$ have been performed in~\cite{Salvio:2023ext,Banerjee:2024cwv}.

\subsubsection{Improved supercool expansion beyond the leading order}\label{improNLO}

So far the improved  supercool expansion has been performed at LO: we have included the cubic term in~(\ref{barVnlo}) in the non-perturbative computation of the bounce action but we have neglected the other corrections. Going beyond the LO in the improved  supercool expansion consists in including the effect of  the other corrections perturbatively. 

 The next step, the improved supercool expansion at NLO, would consist in including  the $x^2\log x$ and $x^2$ terms in the small-$x$ expansion of $J_B(x)$ and $J_F(x)$, see~(\ref{JBdef})-(\ref{JFdef}), and considering the field-dependent logarithm in~(\ref{logSplit}). As already remarked, the $\log a_B$ and $\log a_F$ contributions (and thus the $x^2$ terms) in~(\ref{JBdef})-(\ref{JFdef}) can be ignored because they  can be absorbed in a redefinition of the renormalization-scheme-dependent coefficient $a_s$ in~(\ref{Vqs}).
  In doing these improvements one  needs to include extra parameters beyond $\chi_0$, $\bar \beta$ $g$ and $\tilde g$. An attempt to go to the improved supercool expansion at NLO was performed in a specific model in~\cite{Pascoli:2026tuu}. However, really going to this higher level of precision would also require estimating all functional determinants appearing in the one-loop prefactor of $\Gamma$ given in~\cite{Linde:1980tt,Linde:1981zj}, which were missed in~\cite{Pascoli:2026tuu}. 
So in the improved supercool expansion at NLO one needs to correct Eq.~(\ref{lteq})
for the nucleation temperature $T_n$ in a non-trivial way.
A complete analysis of the improved supercool expansion at NLO is thus not yet present in the literature. 

Moreover, if we were to apply this NLO machinery to GW production one should make sure that the spectrum of GWs used is known at a correspondingly  high level of precision. The inclusion of the $x^2\log x$ terms in the small-$x$ expansion of $J_B(x)$ and $J_F(x)$ in~(\ref{JBdef})-(\ref{JFdef}) improves the approximation if one knows the spectrum of GWs at a level of precision of  $~6/(\pi\times 32)$ (which is a few percents). This number is obtained by comparing the coefficient of the $x^{3/2}$ term in~(\ref{JBdef}) with that of the $x^2\log x$ terms. Achieving this goal is highly non trivial. For example, the templates of the spectrum of GWs used in~\cite{Pascoli:2026tuu} (which were taken from~\cite{Caprini:2024hue}) are affected by  uncertainties whose size is (at least currently) unknown\footnote{The author thanks Germano Nardini, an author of~\cite{Caprini:2024hue}, for a useful private communication on this topic.} and thus  cannot guarantee a few-percent precision. Indeed, in~\cite{Caprini:2024hue} it is explicitly stated that their templates, despite embodying the current state of the art in simplicity and accuracy, are ultimately placeholders for the more precise templates that the community will need to develop over the next decade.

An aspect that can potentially simplify the determination of the spectrum of GWs is that  in the RSB scenario the leading source of GWs is provided by bubble collisions that take place in the vacuum. This is because the energy density of the region through which the bubbles propagate is dominated by the vacuum energy associated with $\chi$, leading to the exponential growth of the cosmological scale factor, as discussed above. This inflationary behavior, as usual, dilutes any preexisting matter and radiation; therefore, one can neglect the production of GWs from hydrodynamic effects such turbulence and sound waves in the cosmic fluid~\cite{Caprini:2015zlo,Maggiore:2018sht,Lewicki:2022pdb}.
  Indeed, one can explicitly check that in our case the condition provided by Ref.~\cite{Caprini:2019egz} for plasma
effects to be negligible is satisfied by using the validity of the supercool expansion (even in its improved form).

\section{Conclusions}\label{Conclusions}
Let us conclude by giving a detailed summary of this review paper.
\begin{itemize}
\item Sec.~\ref{General theoretical framework} summarized  the general theoretical framework of RSB theories, providing the general form of the matter Lagrangian for scalar, vector and fermion fields, reviewing RSB both at zero and finite temperature and showing that RSB always leads to FOPTs. The same section also illustrated how supercooling is a general property of this scenario and showed the validity of perturbation theory to compute the effective action in the relevant regimes. 
\item Sec.~\ref{LO} introduced the supercooling expansion, which relies on the smallness of the parameter $\epsilon$ in~(\ref{CondConv}), focusing on the LO approximation. The tunneling process and the strength $\alpha$ were discussed and the nucleation temperature $T_n$  and the inverse duration $\beta$ were determined. Furthermore, Sec.~\ref{LO} discussed the role of gravity in the vacuum decay and the time-dependence of the corresponding rate in the RSB scenario with large-enough supercooling. 
\item The supercool expansion was then extended beyond the LO in Sec.~\ref{NLO}. The NLO approximation was discussed in some detail and the way even higher-order corrections could be included was briefly illustrated.
\item Sec.~\ref{impro} extended the validity of the supercool expansion to order-one values of $\epsilon$. The supercool expansion of Sec.~\ref{LO} is still valid if there are several degrees of freedom with dominant couplings (all of the same order of magnitude) to the flat-direction field $\chi$. If this is not the case, the validity of the supercool expansion can be extended to order-one values of $\epsilon$ by including the cubic term in $\chi$ in the non-perturbative calculation of the bounce action. The latter approach is referred to as the improved supercool expansion.   The bounce solutions and the corresponding actions were discussed and $T_n$ and $\beta$ were determined at LO in the improved supercool expansion ($\alpha$ is always large in the general RSB scenario). Furthermore, the way higher-order corrections could be included was briefly illustrated.
\end{itemize}

 \vspace{0.7cm}
 
 \subsection*{Acknowledgments}
The author thanks Orlando Luongo and Germano Nardini for valuable discussions and  Francesco Rescigno and Filippo Cutrona for valuable discussions and collaboration on some of the topics of this paper.

\vspace{1cm}

 \footnotesize
\begin{multicols}{2}

\end{multicols}

  \end{document}